\documentclass[american,aps,pra,reprint, superscriptaddress]{revtex4-1}
\usepackage{color}
\usepackage{times}
\usepackage{subfig}
\usepackage{bm}
\usepackage{graphicx}
\usepackage{graphics}
\DeclareGraphicsExtensions{.pdf,.png,.jpg}
\usepackage{amsbsy}
\usepackage{amsmath}
\usepackage{amsfonts}
\usepackage{amsthm}
\usepackage{float}
\usepackage{amssymb}
\usepackage{textcomp}
\usepackage{mathtools}
\usepackage{hyperref}
\usepackage[font=small,labelfont=bf,
justification= centerlast,
format=hang]{caption}

\usepackage[subnum]{cases}
\usepackage{enumitem}
\usepackage[font=small,labelfont=bf,
justification= centerlast,
format=hang]{caption}

\definecolor{purple(html/css)}{rgb}{0.5, 0.0, 0.5}
\newcommand{\ket}[1]{| #1 \rangle}
\newcommand{\bra}[1]{\langle #1 |}
\newcommand{\braket}[2]{\langle #1 | #2 \rangle}

\usepackage{enumerate}
\usepackage{braket}
\usepackage{adjustbox}
\usepackage{footnote}
\usepackage[dvipsnames]{xcolor}
\usepackage{tipa}
\usepackage{textcomp}
\usepackage{fontenc}
\usepackage{mathrsfs}
\usepackage{color}
\usepackage{colortbl}
\usepackage{graphics,graphicx}
\usepackage{amsmath}
\usepackage{bbm}

\begin{document}

\title{Extraction of   Product and Higher Moment Weak Values: Applications in Quantum State Reconstruction and Entanglement Detection}

\author{Sahil}
\email{sahilmd@imsc.res.in}
\affiliation{Optics \& Quantum Information Group, The Institute of Mathematical Sciences, HBNI, CIT Campus, Taramani, Chennai 600113, India.}

\author{Sohail}
\email{sohail@hri.res.in}
\affiliation{Quantum Information \& Computation Group, Harish-Chandra Research Institute, HBNI, Allahabad 211019, India.}

\author{Subhrajit Modak}
\email{modoksuvrojit@gmail.com}
\affiliation{Department of Physical Sciences, Indian Institute of Science Education and Research Mohali, Sector 81, S. A. S. Nagar, Manauli PO 140306, Punjab, India.}

\author{Sibasish Ghosh}
\email{sibasish@imsc.res.in}
\affiliation{Optics \& Quantum Information Group, The Institute of Mathematical Sciences, HBNI, CIT Campus, Taramani, Chennai 600113, India.}

\author{Arun Kumar Pati}
\email{akpati@hri.res.in}
\affiliation{Quantum Information \& Computation Group, Harish-Chandra Research Institute, HBNI, Allahabad 211019, India.}
\begin{abstract}
{ Weak measurements}  introduced by Aharonov, Albert and Vaidman (AAV)   can provide informations about the system with minimal back action.  Weak values of product   observables (commuting)  or higher moments of an observable  are informationally important in the sense that  they are useful to resolve some paradoxes, realize strange quantum effects, reconstruct density matrices, etc.  In this work, we  show that it is possible to access  the higher  moment weak values of an observable  using  weak values of  that  observable with  pairwise  orthogonal post-selections. Although the higher  moment weak values  of an observable are inaccessible  with Gaussian pointer states, our method allows any  pointer state. We have calculated product weak values in a bipartite system for any  given pure and mixed pre selected states. Such product weak values can be obtained using only the measurements of  local weak values (which are defined as single system weak values in a multi-partite system).  As an application, we  use  higher moment weak values and product weak values  to reconstruct  unknown quantum states of  single and bipartite systems, respectively.   Further, we give a necessary separability criteria for finite dimensional systems using  product weak values and  certain class of entangled states violate this inequality  by cleverly choosing the product observables and the post selections. By such choices,  positive partial transpose (PPT) criteria  can be achieved for these classes of entangled states.  Robustness of our method  which occurs  due to inappropriate choices of quantum observables and  noisy post-selections is also discussed here. Our method can easily be generalized to the multi-partite systems.
\end{abstract}
\maketitle
\section{Introduction}
\normalsize {In} quantum systems, measurement of product observables of two or more can contain information  about quantum correlations and quantum dynamics.   In the recent past, many theoretical and experimental works have been performed regarding the  measurements of product observables using weak measurements.   They have been used to resolve the Hardy's Paradox \cite{1} with experimental verification \cite{2}, EPR-Bohm experiment \cite{3}, direct measurement of a density matrix \cite{4}, reconstruction of  entangled quantum states \cite{5}. It has also been reported that  a strange  quantum effect namely the ``Quantum Cheshire Cats"  where the properties of a quantum particle can be disembodied (e.g., photon's position and polarization degrees of freedoms can be separated from each other) can be realized using product weak values \cite{6}. See also the references \cite{7} and \cite{8} for the experimental test of  the existence  of Quantum Cheshire Cats and the exchange of  grins between two such Quantum Cheshire Cats, respectively.  Weak measurements of  product observables also play an important role in  understanding the quantum mechanics  such as Bell tests \cite{9,10,11}, nonlocality via post-selection \cite{12}.\par 
Higher moment weak values are useful to obtain the weak-valued probability distribution \cite{13} of an observable in the pre and post-selected systems (which will be shown in this work later).  Some applications  of  weak-valued probability distribution  are: (a) Ozawa’s measurement- disturbance relation has experimentally been verified using weak-valued probability distribution \cite{14}, (b) to obtain all of the values of the observables relevant to a Bell test experimentally, weak-valued probabilities have been used \cite{15}, (c) the authors of \cite{16} have shown that there is a connection between weak-valued joint probabilities and incompatibility. There are some other applications of weak-valued probabilities e.g., (d) experimental realization of the Quantum Box Problem \cite{17}, (e) to resolve Hardy’s paradox \cite{2}, (f) justification of Scully \emph{et al.}’s claim \cite{18} that the momentum disturbance associated with which-way measurement in Young’s double-slit experiment can be avoided has been shown by the negativity of the weak-valued probabilities corresponding to the momentum disturbance, which consequently have zero variance \cite{19,20}, (g) to control the probe wave packet of the target system by pre and post-selections, one can use the higher moment weak values \cite{21}, (h) to obtain the modular value of an observable in pre and post- selected systems, higher moment weak values can be used as there is an exact connection between them \cite{22,23}.
\par
Weak measurements in particular weak values are known to provide useful informations with simple experimental setups. The  ``weak value" was first  introduced by  Aharonov, Albert, and Vaidman (AAV)  \cite{24}. It  was inspired by the two-time formulation of the quantum-mechanical system \cite{25}.  The mechanism of  AAV method is that they took the von Neumann measurement scheme  \cite{26} one step forward  by considering the coupling coefficient very small i.e., weak followed by a strong measurement in succession. This formulation is characterized by the pre- and postselected states of the system. By preparing  a system initially  in the state $\ket{\psi}$ and  post-selecting  in the state $\ket{\phi}$, as a result we obtain the weak value of any observable ${A}$ which  is defined as 
\begin{align}
{\braket{{A}_w}}^{\phi}_{\psi}=\frac{\braket{\phi |{A}|\psi}}{\braket{\phi |\psi}}\label{1}.
\end{align}
This is a complex number and the spatial and momentum displacements of the pointer state give the real and imaginary  parts of that weak value \cite{27,28}, respectively. By this way, we get the full knowledge of the complex weak value. One of the exciting and interesting features of weak value is that it can lie outside the max-min range of the eigenvalues of the operator of interest. Simultaneously, measurement disturbance is quite small which gives one to perform further measurements or simultaneous measurement of multiple observables. \par
Weak measurements have been proven useful in understanding quantum systems such as  for the direct measurement of the wave function of a quantum system \cite{29},  to calculate slow- and fast-light effects in birefringent photonic crystals \cite{30}, the confirmation of the Heisenberg-Ozawa uncertainty relationship \cite{31}, for detecting tiny spatial shifts \cite{32,33}.  Weak value can also  be used to measure non-Hermitian operators \cite{34,35}, in hot thermometry \cite{36}, to detect entanglement universally in a two-qubit system \cite{37}. \par
\normalsize{\textit{Product weak values:}} The observables used in the von Neumann measurement scheme  are of simple kinds. Direct measurement of product observables is extremely  difficult whether it is strong or weak. This difficulty arises from the fact that the measurement interaction in the von Neumann measurement scheme couples two different observables to a single pointer \cite{27}. To overcome this, an approach using multi particle interaction Hamiltonian was proposed by Resch and Steinberg (RS) {\cite{38}} applying AAV method. Namely,  they have used the Hamiltonian of the form ${H}=g_1{A}\otimes{p}_x + g_2{B}\otimes{p}_y$ with gaussian pointer states. Here $A$ and $B$ are two observables of the system. ${p}_x$ and ${p}_y$ are the pointer's momentum degrees of freedom in two different directions x and y, respectively. $g_1$ and $g_2$ are coupling coefficients between the system and the pointer for  two different directions, respectively.  By performing a second order expansion in the two-dimensional pointer displacement $\braket{{X} {Y}}$, they showed that it is possible to extract the real part of the product weak value  for the case of commuting observables $[{A},{B}]=0$, namely $Re[\braket{(AB)_w}_{\psi}^{\phi}]=2{\braket{XY}}/({g_1g_2t^2})-Re[(\braket{A_w}_{\psi}^{\phi})^* \braket{B_w}_{\psi}^{\phi}]$, where `t' is the interaction time.  For imaginary part of the product weak value, one needs to look into $\braket{{X} {P}_y}$.  Note that, the weak value of a tensor product  observable  in a bipartite system (we will call it  product weak value in a bipartite system)  can also be calculated according to the RS method as the local observables are commuting. It's generalised version i.e., the product  weak values $\braket{({A} {B})w}_{\psi}^{\phi}$ and it's higher orders i.e., $\braket{({A} {B} {C}\cdots)w}_{\psi}^{\phi}$ using N pointers’ correlations  can be found in Ref. {\cite{39}}.\par
\normalsize{\textit{Summary of the present work:}} In this work, we show that by introducing  an unique  orthogonal  state to the given post selection, it is possible to extract the higher moment weak values $\braket{({A}^n)_w}_{\psi}^{\phi}$ in a single system  as well as the product weak values for the given  pure and mixed pre selected  states in a bipartite system separately.  We   show application of our results to  reconstruct quantum states of  single and bipartite systems. We have used higher moment weak values to reconstruct a  pure state of a single system and product weak values of  a bipartite system to reconstruct unknown pure and mixed states of that  bipartite system. Recently, Pan \emph{et al.} \cite{5} have used product weak values of projection operators  to reconstruct an unknown bipartite pure state. They have considered entangled pointer states as well as the modular values of  local projection operators  and the modular values of  sum of the local  projection operators \cite{23} to evaluate product weak values. We, for the first time show that such product weak values can be realized locally for the case of both pure and mixed states. Also, to reconstruct the states of single and bipartite systems,  we have generalized the  measurement of projection operators to  arbitrary  observables. Our method can be generalized to the multipartite systems.  Further, we give a necessary separability inequality for finite dimensional systems using  product weak values. This inequality is violated by  certain class of entangled states  by cleverly choosing the product observables and the post selections. By such choices, The PPT criteria can be achieved for these class of entangled states. In particular, we give several  examples namely $(i)$ two-qubit Werner state (noisy singlet), $(ii)$ mixture of two-Bell states, $(iii)$ mixture of arbitrary pure entangled and maximally mixed states, $(iv)$ mixture of two arbitrary entangled states, $(v)$ mixture of four-Bell states, $(vi)$ two qudit  Werner states, $(vii)$ higher dimensional isotropic states. The criteria can potentially detect more classes of entangled states with suitably choosing product observables and post-selections. Finally we show that our methods of  “extraction of product  and higher moment weak values”  are robust against the errors which occur  due to the inappropriate choices of system observables and unsharp post-selections.\par
This paper is organized as follows. In sec. \ref{e II} we provide the formulation of our method.  In sec. \ref{e III} we apply our method to reconstruct quantum states of single as well as bipartite systems separately. Entanglement detection criteria is shown in sec. \ref{e IV}. We show the robustness of our method in sec. \ref{e V} and finally conclude in sec. \ref{e VI}.
\section{FORMULATION}\label{e II}
The following identity which is sometimes referred as Vaidman’s formula \cite{40}  will be used to derive the main results of this paper
\begin{align}
{A}\ket{\phi}=\braket{{A}}_{\phi}\ket{\phi} + \braket{\Delta A}_{\phi} \ket{\phi^{\perp}_{A}},\label{2}
\end{align}
where $A$ is an Hermitian operator and $\ket{\phi}$ is any quantum state vector in the Hilbert space $\mathcal{H}$. The state vector  $\ket{\phi_{A}^{\perp}}$ is orthogonal to  $\ket{\phi}$, $\braket{{A}}_{\phi}=\braket{\phi |{A}|\phi}$ and $\braket{\Delta A}_{\phi}=\braket{\phi^{\perp}_{A} |{A}|\phi}$. For the derivation see Appendix \ref{A}
\subsection{Higher moment weak values}\label{II A}
If $\ket{\psi}$, ${A}$ and $\ket{\phi}$ are the pre-selected state, the observable and the post-selected state, respectively, then the weak value of the observable $A$ is given by
\begin{align}
{\braket{{A}_w}}^{\phi}_{\psi}&=\left(\frac{\braket{\psi|{A}|\phi}}{\braket{\psi|\phi}}\right)^{\ast}\!=\braket{A}_{\phi} +\braket{\Delta A}_{\phi}\frac{\braket{\phi_{A}^\perp|\psi}}{\braket{\phi|\psi}},\label{3}
\end{align}
where we have used Eq. (\ref{2}).  A  similar expression was considered in Ref. \cite{41} to explain the origin of the complex  and anomalous nature of a weak value.  The Eq. (\ref{3})  for the expression of the weak value will be useful for deriving the following result.\par
\emph{Result 1:-}
The weak value of the operator $A^2$ which we call  the ``second moment weak value" has the following expression
 \begin{align}
{\braket{({A}^2)_w}}^{\phi}_{\psi}=\braket{A}_\phi \Big({\braket{{A}_w}}^{\phi}_{\psi} - {\braket{{A}_w}}^{\phi^{\perp}_{A}}_{\psi}\Big) + {\braket{{A}_w}}^{\phi}_{\psi}{\braket{{A}_w}}^{\phi^{\perp}_{A}}_{\psi},\label{4}
\end{align}
where ${\braket{{A}_w}}^{\phi}_{\psi}$ and ${\braket{{A}_w}}^{\phi^{\perp}_{A}}_{\psi}$ are the weak values of the operator $A$ for the given pre selection $\ket{\psi}$ with two post-selections $\ket{\phi}$ and $\ket{{\phi^{\perp}_{A}}}$, respectively.
\begin{proof}
\begin{align}
{\braket{({A}^2)_w}}^{\phi}_{\psi}&=\left(\frac{\braket{\psi |{A}^2|\phi}}{\braket{\psi|\phi}}\right)^{\ast}\nonumber\\
&=\left(\frac{1}{\braket{\psi|\phi}}\bra{\psi}{A}[\braket{A}_{\phi}\ket{\phi} +\braket{\Delta A}_{\phi}\ket{\phi_{A}^\perp}]\right)^\ast \nonumber\\
&=\left(\braket{A}_{\phi}\frac{\braket{\psi |{A}|\phi}}{\braket{\psi|\phi}} +\braket{\Delta A}_{\phi}\frac{\braket{\psi |{A}|\phi_{A}^\perp}}{\braket{\psi|\phi}}\right)^\ast \nonumber\\
&=\braket{A}_{\phi}\frac{\braket{\phi |{A}|\psi}}{\braket{\phi|\psi}} +\braket{\Delta A}_{\phi}\frac{\braket{\phi_{A}^\perp |{A}|\psi}}{\braket{\phi_{A}^\perp |\psi}}\frac{\braket{\phi_{A}^\perp |\psi}}{\braket{\phi|\psi}}.\label{5}
\end{align}
From Eq. (\ref{3}), using $\braket{\Delta A}_{\phi}\frac{\braket{\phi_{A}^\perp|\psi}}{\braket{\phi|\psi}}=\braket{A_w}^{\phi}_{\psi}-\braket{A}_{\phi} $ in Eq. (\ref{5}), we will obtain Eq. (\ref{4}).
\end{proof}
In the similar way we obtain all the higher moment weak values which take the general form as
\begin{align}
{\braket{{A}^n_w}}^{\phi}_{\psi}=\braket{A}_\phi \hspace{-2pt}\Big(\hspace{-3pt}{\braket{{A}^{n-1}_w}}^{\phi}_{\psi}\! - \braket{{A}^{n-1}_w}^{\phi^{\perp}_{A}}_{\psi}\hspace{-2pt}\Big) \!+ {\braket{{A}^{n-1}_w}}^{\phi^{\perp}_{A}}_{\psi}{\braket{{A}_w}}^{\phi}_{\psi},\label{6}
\end{align}
for $n=1,2,\cdots$. \par
Now consider the second moment weak value i.e., Eq. (\ref{4}), where ${\braket{{A}_w}}^{\phi}_{\psi}$ and ${\braket{{A}_w}}^{\phi^{\perp}_{A}}_{\psi}$ are extractable from the one and the same experimental set-up for the post-selection of $\ket{\phi}$ and $\ket{\phi^{\perp}_{A}}$, respectively  as these two states are orthogonal to each other. Note that, although the post-selection can be realized here in one and the same measurement set-up, nevertheless, in order to actually find out the weak values ${\braket{{A}_w}}^{\phi}_{\psi}$ and ${\braket{{A}_w}}^{\phi^{\perp}_{A}}_{\psi}$, measurements of phase-space displacements for the two post-selected states $\ket{\phi}$ and $\ket{\phi^{\perp}_{A}}$  need to be performed. $\braket{A}_\phi$ is the average value of $A$ for the post selected state $\ket{\phi}$. Extraction of higher moment weak values becomes extremely simple in two dimensional Hilbert space  discussed in the following.\par
{\it  Two dimensional case:} In two dimensional Hilbert space,  there are only two pairwise orthogonal post-selections which occur at the same time and hence both the weak values with orthogonal post-selections can be extracted simultaneously. Particularly in this dimension, we find that only by knowing the weak  values ${\braket{{A}_w}}^{\phi}_{\psi}$ and ${\braket{{A}_w}}^{\phi^{\perp}_{A}}_{\psi}$,  we are able to obtain all the higher moment weak values without any further complications in comparison to Ref.  \cite{42}. So the  number of measurements is being reduced considerably than the earlier proposal \cite{42}. \par
{\it  Higher dimensional case:} In higher dimensional Hilbert space, to extract second moment weak value of the observable $A$, one can  perform  the projective measurements $\{\ket{\phi}\bra{\phi}, \ket{\phi^{\perp}_{A}}\bra{\phi^{\perp}_{A}}, I-\ket{\phi}\bra{\phi} -\ket{\phi^{\perp}_{A}}\bra{\phi^{\perp}_{A}} \}$ for post-selections.\par
It can be shown that  the $n$-th moment weak value i.e., ${\braket{A^{n}_w}}^{\phi}_{\psi}$, consists  $n$ number of different weak values, namely  ${\braket{{A}_w}}^{\phi}_{\psi}$, ${\braket{{A}_w}}^{\phi^{\perp}_{A}}_{\psi}$,$\cdots$,${\braket{{A}_w}}^{(\phi^{\perp}_{A})^{n-1}}_{\psi}$.  Here  $\ket{(\phi^{\perp}_{A})^{n-1}}=\ket{((\phi^{\perp}_{A})^{\perp}_{A})\cdots (n-1) \hspace{1mm} times}=\frac{1}{\braket{\Delta A}_{(\phi^{\perp}_{A})^{n-2}}}\left(A-\braket{A}_{(\phi^{\perp}_{A})^{n-2}}\right)\ket{(\phi^{\perp}_{A})^{n-2}}$. Now, if $n$ is even, then there exist pairwise orthogonal  post-selected states i.e.,
$\left(\ket{\phi},\ket{\phi^{\perp}_{A}}\right)$, $\left(\ket{(\phi^{\perp}_{A})^{\perp}_{A}},\ket{((\phi^{\perp}_{A})^{\perp}_{A})^{\perp}_{A}}\right)$,$\cdots$,$\left(\ket{(\phi^{\perp}_{A})^{n-2}},\ket{(\phi^{\perp}_{A})^{n-1}}\right)$. So, it is possible to obtain weak values  ${\braket{{A}_w}}^{\phi}_{\psi}$ and  ${\braket{{A}_w}}^{\phi^{\perp}_{A}}_{\psi}$ simultaneously for the first pair of post-selected states, so on and so forth. So, effectively the total  number of measurements to be performed according to the AAV method to extract the $n$-th moment weak value is ${n}/{2}$. For odd $\hspace{0.5mm}$ $n$, the number of measurements is $(n+1)/2$. Note that, all the measurements here are to be done by only  changing the post-selections while keeping  the observable $A$  fixed in the AAV method.  \textit{Once we extract the $n$-th moment weak value, then all the lower moment weak values can be calculated  from the data of $n$-th moment weak value.}\par 
The higher moment weak values of an observable are inaccessible with Gaussian pointer states. The reason is that, in the RS method \cite{38}, the higher moment weak value terms will vanish due to the  properties of the Gaussian  pointer state. The whole expression can be found in Ref. \cite{42} (equation 4).  More specifically, it can be seen in the above mentioned expression that when the orbital angular momentum (OAM) of the pointer state is zero (which corresponds to the two dimensional Gaussian pointer state, i.e., OAM state with zero orbital angular momentum), the higher moment weak value terms vanish. To retrieve higher moment weak values, we need to use OAM states with non-zero orbital angular momentums. The key factor for such cases is that the two dimensional OAM states are not factorizable in two different directions for non-zero orbital angular momentums. In doing that one needs to engineer OAM states with higher winding numbers, or superpositions of OAM states to obtain  the higher moment weak values. This procedure  can become difficult as the moments increase. One needs to prepare pointer states with different combination of orbital angular momentum {\cite{42}}. Moreover, there are several disadvantages of the RS method from experimental perspective which we will discuss later in this section. See \cite{43} for a comment regarding the extraction of higher moment weak values. \par
As an application, we will use the higher moment weak values to reconstruct an unknown pure state of a single system. See  Appendix \ref{B} for the derivation of the second or higher moment weak values  for the mixed pre selected state case.
\subsection{Product weak values}{}\label{II B}
\emph{Product weak values for pure pre-selected state:}
The product  weak value of the  observable  ${A}\otimes {B}$ in a bipartite system  is given by
\begin{align}
  \braket{({A}\otimes {B})_w}^{\phi_A\phi_B}_{\psi_{AB}}=\frac{\braket{\phi_A\phi_B|({A}\otimes {B})|\psi_{AB}}}{\braket{\phi_A\phi_B|\psi_{AB}}},\label{7}
\end{align}
 where the pre selection  is the bipartite pure state $\ket{\psi_{AB}}$ and the post-selection is a product state $\ket{\phi_A\phi_B}=\ket{\phi_A}\otimes \ket{\phi_B}$.\par
Consider the ``local weak value" of the operator $A$
 \begin{align}
  \braket{A^{local}_w}^{\phi_A\phi_B}_{\psi_{AB}}&=\frac{\braket{\phi_A\phi_B|({A}\otimes {I_B})|\psi_{AB}}}{\braket{\phi_A\phi_B|\psi_{AB}}}\nonumber\\
  &=\braket{A}_{\phi_A} + \braket{\Delta A}_{\phi_A}\frac{\braket{\phi^{\perp}_A\phi_B|\psi_{AB}}}{\braket{\phi_A\phi_B|\psi_{AB}}},\label{8}
\end{align}
where we have used Eq. (\ref{2}) in the subsystem A. Eq. (\ref{8}) will be used to derive the following result.
 \begin{widetext}
 \emph{Result 2:-} The product weak value in Eq. (\ref{7}) can be realized via local weak values as
\begin{align}
\braket{(A\otimes B)_w}^{\phi_A\phi_B}_{\psi_{AB}}=&\braket{A}_{\phi_A} \left({\braket{B^{local}_w}}^{\phi_A\phi_B}_{\psi_{AB}} - {\braket{{B^{local}_w}}^{\phi^{\perp}_{A}\phi_B}_{\psi_{AB}}}\right) + {\braket{A^{local}_w}}^{\phi_A\phi_B}_{\psi_{AB}}{\braket{B^{local}_w}}^{\phi^{\perp}_{A}\phi_B}_{\psi_{AB}}\label{9}
\end{align}
 where ${\braket{{A}^{local}_w}}^{\phi_A\phi_B}_{\psi_{AB}}$, ${\braket{{B}^{local}_w}}^{\phi_A\phi_B}_{\psi_{AB}}$ and ${\braket{{B}^{local}_w}}^{\phi^{\perp}_{A}\phi_B}_{\psi_{AB}}$ are the ``local'' weak values and  $\ket{\phi^{\perp}_{A}}=\frac{1}{\braket{\Delta A}_{\phi_A}}\left(A - \braket{A}_{\phi_A}\right)\ket{\phi_A}$ is given by  (\ref{2}) for the subsystem A.
 \begin{proof}
\begin{align}
\braket{({A}\otimes {B})_w}^{\phi_A\phi_B}_{\psi_{AB}}&=\left(\frac{\braket{\psi_{AB}|({A}\otimes {B})|\phi_A\phi_B}}{\braket{\psi_{AB}|\phi_A\phi_B}}\right)^\ast\nonumber\\
&=\left(\frac{\braket{\psi_{AB}|(\braket{A}_{\phi_A}\ket{\phi_A} +\braket{\Delta A}_{\phi_A}\ket{\phi_{A}^\perp})\otimes {B}|\phi_B}}{\braket{\psi_{AB}|\phi_A\phi_B}}\right)^\ast\nonumber\\
&=\braket{A}_{\phi_A}\frac{\braket{\phi_A\phi_B|({I_A}\otimes {B})|\phi_{AB}}}{\braket{\phi_A\phi_B|\psi_{AB}}} + \braket{\Delta A}_{\phi_A}\frac{\braket{\phi^{\perp}_A\phi_B|({I_A}\otimes {B})|\phi_{AB}}}{\braket{\phi^{\perp}_A\phi_B|\psi_{AB}}}\frac{\braket{\phi^{\perp}_A\phi_B|\psi_{AB}}}{\braket{\phi_A\phi_B|\psi_{AB}}}\nonumber\\
&=\braket{A}_{\phi_A}{\braket{{B}^{local}_w}}^{\phi_A\phi_B}_{\psi_{AB}} + {\braket{{ B}^{local}_w}}^{\phi^{\perp}_A\phi_B}_{\psi_{AB}}\left({\braket{{A}^{local}_w}}^{\phi_A\phi_B}_{\psi_{AB}} - \braket{A}_{\phi_A}\right),\nonumber
\end{align}
where we have  used Eq. (\ref{2}) in the second line for the subsystem A and Eq. (\ref{8}) in the fourth line. After the manipulation, we have Eq. (\ref{9}).
\end{proof}
 \end{widetext}\par
\emph{Product weak values in terms of local weak values:}  We have obtained a product weak value using only local weak values. Note that the local weak values ${\braket{{B}^{local}_w}}^{\phi_A\phi_B}_{\psi_{AB}}$ and ${\braket{{ B}^{local}_w}}^{\phi^{\perp}_{A}\phi_B}_{\psi_{AB}}$ can be measured in the same experimental setup as the post-selected states $\ket{\phi_A}$ and $\ket{\phi^{\perp}_{A}}$ are orthogonal to each other. We need another measurement setup for ${\braket{{A}^{local}_w}}^{\phi_A\phi_B}_{\psi_{AB}}$. So, effectively the total number of  measurements  to be performed according to the AAV method to extract the product weak value $\braket{({A\otimes B})_w}^{\phi_A\phi_B}_{\psi_{AB}}$ is only two.  In experiment, local weak values like ${\braket{{B}^{local}_w}}^{\phi_A\phi_B}_{\psi_{AB}}$ can be realized as 
\begin{align}
  \braket{B^{local}_w}^{\phi_A\phi_B}_{\psi_{AB}}&=\frac{\braket{\phi_B\phi_A|(I_{A}\otimes {B})|\psi_{AB}}}{\braket{\phi_B\phi_A|\psi_{AB}}}\label{10}\\
  &=\frac{\braket{\phi_B|{B}|\psi^{\phi_A}_B}}{\braket{\phi_B|\psi_B^{\phi_A}}},\label{11}
\end{align}
where $\ket{\psi_B^{\phi_A}}=\braket{\phi_A|\psi_{AB}}$ is an unnormalized state for the  subsystem B. That is, we first measure the projection operator $\Pi_{\phi_A}=\ket{\phi_A}\bra{\phi_A}$ on the subsystem A, then the  state of the subsystem B becomes $\braket{\phi_A|\psi_{AB}}/\sqrt{\braket{\psi_B^{\phi_A}|\psi_B^{\phi_A}}}$ which we consider to be pre-selected state for the subsystem B. This pre-selected state is unknown as $\ket{\psi_{AB}}$ is unkown. So, for each given projector related to  the subsystem A, there exists an unknown pre-selected state  of the subsystem B. The observable is $B$ and the post-selection is $\ket{\phi_B}$. So from Eq. (\ref{11}), we  see that the local weak value $ \braket{B^{local}_w}^{\phi_A\phi_B}_{\psi_{AB}}$ can be realized  according to the AAV method related to the subsystem B. \par
\emph{Product weak values for mixed pre-selected state:} If the pre selection of the bipartite system is a mixed state $\rho_{AB}$ then, the product weak value of the observable ${A}\otimes {B}$ is given by 
 \begin{align}
  \braket{({A}\otimes {B})_w}^{\phi_A\phi_B}_{\rho_{AB}}=\frac{\braket{\phi_A\phi_B|\left({A}\otimes {B}\right)\rho_{AB}|\phi_A\phi_B}}{\braket{\phi_A\phi_B|\rho_{AB}|\phi_A\phi_B}}.\label{12}
\end{align}
Note that generalization for mixed state of  Eq. (\ref{9}) is not straightforward.
  \begin{widetext}
 \emph{Result 3:-} The product weak value in Eq. (\ref{12}) can be realized via local weak values as
 \begin{align}
\braket{(A\otimes B)_w}^{\phi_A\phi_B}_{\rho_{AB}}=\braket{A}_{\phi_A}\braket{B^{local}_w}^{\phi_A\phi_B}_{\rho_{AB}} + \frac{\braket{\Delta A}_{\phi_A}}{2p(\rho_{AB},\phi_A\phi_B)}\sum_{i=1}^{m}\bigg( &\lambda^i_A \braket{ B^{local}_w}^{i_A\phi_B}_{\rho_{AB}}p(\rho_{AB},i_A\phi_B) \nonumber\\
&+ {\lambda^{\prime}_A}^i \braket{ B^{local}_w}^{{i^{\prime}_A}\phi_B}_{\rho_{AB}}p(\rho_{AB},{i^{\prime}_A}\phi_B) \bigg),\label{13}
\end{align}
where \{$\lambda_A^i,\ket{i_A}$\} and \{${\lambda^{\prime}_A}^i,\ket{{i^{\prime}_A}}$\} satisfy the spectral decomposition for the  normal operators $\ket{\phi_A}\bra{\phi^{\perp}_{A}} + \ket{\phi^{\perp}_A}\bra{\phi_A}$ and $\ket{\phi_A}\bra{\phi^{\perp}_{A}} - \ket{\phi^{\perp}_A}\bra{\phi_A}$, respectively. $p(\rho_{AB},\phi_A\phi_B)=\braket{\phi_A\phi_B|\rho_{AB}|\phi_A\phi_B}$  is the probability of obtaining the post-selected state $\ket{\phi_A\phi_B}$ for  the given  pre selected state  $\rho_{AB}$ and `$m$' is the  dimension of the  subsystem A.\par
 \begin{proof}
 \begin{align}
  \braket{({A}\otimes {B})_w}^{\phi_A\phi_B}_{\rho_{AB}}&=\frac{\braket{\phi_A\phi_B|\left({A}\otimes {B}\right)\rho_{AB}|\phi_A\phi_B}}{\braket{\phi_A\phi_B|\rho_{AB}|\phi_A\phi_B}}\nonumber\\
  &=\braket{A}_{\phi_A}\frac{\braket{\phi_A\phi_B|\left({I_A}\otimes {B}\right)\rho_{AB}|\phi_A\phi_B}}{\braket{\phi_A\phi_B|\rho_{AB}|\phi_A\phi_B}} + \braket{\Delta A}_{\phi_A}\frac{\braket{\phi^{\perp}_A\phi_B|\left({I_A}\otimes {B}\right)\rho_{AB}|\phi_A\phi_B}}{\braket{\phi_A\phi_B|\rho_{AB}|\phi_A\phi_B}},\label{14}
\end{align}
where we have used Eq. (\ref{2}). Now $\braket{\phi^{\perp}_A\phi_B|\left({I_A}\otimes {B}\right)\rho_{AB}|\phi_A\phi_B}=Tr\left[\left(\ket{\phi_A}\bra{\phi^{\perp}_{A}}\otimes \ket{\phi_B}\bra{\phi_{B}}\right)(I_A\otimes B)\rho_{AB}\right]$ can be calculated as 
\begin{align}
\braket{\phi^{\perp}_A\phi_B|\left({I_A}\!\otimes \!{B}\right)\rho_{AB}|\phi_A\phi_B} + \braket{\phi_A\phi_B|\hspace{-2pt}\left({I_A}\!\otimes\! {B}\right)\rho_{AB}|\phi^{\perp}_A\phi_B}\hspace{-2pt}=\hspace{-2pt}Tr\hspace{-2pt}\left[\left(\{\ket{\phi_A}\bra{\phi^{\perp}_{A}} + \ket{\phi^{\perp}_A}\bra{\phi_{A}}\}\otimes \ket{\phi_B}\bra{\phi_{B}}\right)\hspace{-2pt}\left(I_A\otimes B\right)\hspace{-2pt}\rho_{AB}\right],\label{15}
\end{align}
where $\ket{\phi_A}\bra{\phi^{\perp}_{A}} + \ket{\phi^{\perp}_A}\bra{\phi_{A}}$ is a normal operator satisfying $XX^{\dagger}=X^{\dagger}X$ (where X is any operator)  and hence can be  written in spectral decomposition 
\begin{align}
\ket{\phi_A}\bra{\phi^{\perp}_{A}} + \ket{\phi^{\perp}_A}\bra{\phi_A}=\sum_i^{m}{\lambda_A^i \ket{i_A}\bra{i_A}},\label{16}
\end{align}
where \{$\ket{i_A}$\} is the set of  eigenvectors with eigenvalues \{$\lambda_A^i$\}. Using Eq. (\ref{16}) in (\ref{15}), we have
\begin{align}
\braket{\phi^{\perp}_A\phi_B|\left({I_A}\otimes {B}\right)\rho_{AB}|\phi_A\phi_B} + \braket{\phi_A\phi_B|\left({I_A}\otimes {B}\right)\rho_{AB}|\phi^{\perp}_A\phi_B}&=\sum_i^{m}{\lambda_A^i\braket{i_A\phi_B|(I_B\otimes B)\rho_{AB}|i_A\phi_B}}\nonumber\\
&=\sum_i^{m}{\lambda_A^i\braket{B^{local}_w}^{i_A\phi_B}_{\rho_{AB}}p(\rho_{AB},i_A\phi_B)},\label{17}
\end{align}
where we have used the definition of  the weak value   $\braket{ B_w}^{i_A\phi_B}_{\rho_{AB}}$ of the observable $B$  for the given pre and post-selections $\rho_{AB}$ and $\ket{i_A\phi_B}$, respectively.  $p(\rho_{AB},i_A\phi_B)=\braket{i_A\phi_B|\rho_{AB}|i_A\phi_B}$ is the probability of obtaining the product state $\ket{i_A\phi_B}$. Now similarly,
\begin{align}
\braket{\phi^{\perp}_A\phi_B|\left({I_A}\otimes {B}\right)\rho_{AB}|\phi_A\phi_B} - \braket{\phi_A\phi_B|\left({I_A}\otimes {B}\right)\rho_{AB}|\phi^{\perp}_A\phi_B}=\sum_i^{m}{{\lambda^{\prime}_A}^i\braket{B^{local}_w}^{i^{\prime}_A}_{\rho_{AB}}p(\rho_{AB},{i^{\prime}_A}\phi_B)},\label{18}
\end{align}
where we have used the fact that $\ket{\phi_A}\bra{\phi^{\perp}_{A}} - \ket{\phi^{\perp}_A}\bra{\phi_A}$ is also a normal operator with the spectral decomposition 
\begin{align}
\ket{\phi_A}\bra{\phi^{\perp}_{A}} - \ket{\phi^{\perp}_A}\bra{\phi_A}=\sum_i^{m}{{\lambda^{\prime}_A}^i \ket{i^{\prime}_A}\bra{{i^{\prime}_A}}}.\label{19}
\end{align}
Now adding two equations (\ref{17}) and (\ref{18}), we have 
\begin{align}
\braket{\phi^{\perp}_A\phi_B|\left({I_A}\otimes {B}\right)\rho_{AB}|\phi_A\phi_B}=\frac{1}{2}\sum_i^{m}{\left(\lambda_A^i\braket{ B^{local}_w}^{i_A\phi_B}_{\rho_{AB}}p(\rho_{AB},i_A\phi_B) + {\lambda^{\prime}_A}^i \braket{B^{local}_w}^{i^{\prime}_A\phi_B}_{\rho_{AB}}p(\rho_{AB},{i^{\prime}_A}\phi_B)\right)}.\label{20}
\end{align}
Using Eq. (\ref{20}) in (\ref{14}), we obtain Eq. (\ref{13}).
\end{proof}
\end{widetext}\par
\emph{Product weak values in terms of local weak values:}  The product weak value $\braket{(A\otimes B)_w}^{\phi_A\phi_B}_{\rho_{AB}}$ consists the local weak values $\braket{ B^{local}_w}^{\phi_A\phi_B}_{\rho_{AB}}$,  $\{\braket{B^{local}_w}^{i_A\phi_B}_{\rho_{AB}}\}^{m}_{i=1}$ and $\{\braket{B^{local}_w}^{i^{\prime}_A\phi_B}_{\rho_{AB}}\}^{m}_{i=1}$. Here `$m$' is the dimension of the subsystem A. Note that, $\{\braket{B^{local}_w}^{i_A\phi_B}_{\rho_{AB}}\}^{m}_{i=1}$ can be measured in the same experimental setup according to the AAV method as the post-selections $\{\ket{i_A}\}$ form complete set of orthogonal basis. Similarly  $\{\braket{B^{local}_w}^{{i^{\prime}_A}\phi_B}_{\rho_{AB}}\}^{m}_{i=1}$ can also be measured within the same experiment with the complete set of post-selections $\{\ket{i^{\prime}_A}\}$ according to the AAV method. \emph{So, the total number of measurements to be performed according to the AAV method to extract  $\braket{(A\otimes B)_w}^{\phi_A\phi_B}_{\rho_{AB}}$ is  only three}. Local weak values can be realized in the same way as discussed above in Eq. (\ref{11}).\par
\emph{Comparison with previous works:} In the Ref. {\cite{27,38}}, authors have shown that it is also possible to extract the product weak values  by obtaining  local weak values of the observables separately as well as by looking into different  pointers' position and momentum correlations i.e., statistical averages of these different degrees of freedom. There are some cases where polarization degrees of freedom (polarization correlations) was considered instead of position and momentum variables \cite{2}. For such cases statistical averages of those polarization degrees of freedom can be difficult to obtain or number of measurements will be large. Instead, our approach shows that  one needs to obtain only local weak values. Our approach is not necessarily  restricted with Gaussian pointer states. But in most of the previous works \cite{27,38,46}, different type  of pointer states have been used with certain constraints unlike in our case. \par 
Our methods are  easy to perform in experiments due to their local realizations. While in the previous works  \cite{27,38,42,46} i.e., extraction of higher moment weak values and product weak values, there are N pointers' correlations to be measured. So the scalability of their methods face significant challenges. Moreover, from an experimental perspective, their schemes are hard to implement as the resources required to observe the correlations are of second order in terms  of the interaction coefficient.\par
In the next section, we discuss about the applications of our results to reconstruct unknown quantum states in a single and bipartite systems.
\section{Quantum State Tomography} \label{e III}
In the following,  we discuss some methods of  quantum state tomography of a single and bipartite system using  higher moments weak values and product weak values.
{\subsection{ State reconstruction of a single system}} 
\subsubsection{Pure state}
As an application of higher moment weak values we reconstruct a  pure state. The method of  quantum state reconstruction using weak values was introduced by Lundeen \emph{et al}. \cite{lundeennature} as follows. Any  state can be written in computational basis $\{\ket{i}\}$ as 
\begin{align}
\ket{\psi}=\sum_{i=0}{\alpha_i\ket{i}},\label{21}
\end{align}
where $\alpha_i=\braket{i|\psi}$. Now the weak value of a projection operator $\Pi_{i}=\ket{i}\bra{i}$ is given by 
\begin{align}
\braket{(\Pi_{i})_w}_{\psi}^{b}=\frac{\braket{b|{i}}\braket{i|\psi}}{\braket{b|\psi}},\label{22}
\end{align}
with $\braket{b|i}\neq0$. where $\ket{b}$ is a post-selection. So using Eq. (\ref{22}), we finally construct the pure state Eq. (\ref{21})
\begin{align}
\ket{\psi}=\sum_i{\frac{\braket{(\Pi_{i})_w}_{\psi}^{b}}{\braket{b|{i}}}\ket{i}}.\label{23}
\end{align}
The complex number $\braket{b|\psi}$ is not taken into account as it corresponds to the global phase factor after normalization. So to measure a  pure state, we need to obtain weak values of the projection operators $\Pi_i$ with pre selection $\ket{\psi}$ and post-selection $\ket{b}$, respectively.\par
Instead of measuring weak values of the  projection operators individually, we want to use the higher moment weak values of the observable which satisfies spectral decomposition with those projection operators and using those higher moment weak values we will obtain weak values of the projection operators. Let the observable  be  
\begin{align}
A=\sum_i{a_i\Pi_{i}},\label{24}
\end{align}
where $a_i$ are the eigenvalues of the observable $A$. Now the weak value and higher moment weak values of the observable are 
 \begin{align}
\braket{A_w}_{\psi}^{b}=\sum_i{a_i\braket{(\Pi_{i})_w}_{\psi}^{b}},\label{25}\\
\braket{A^n_w}_{\psi}^{b}=\sum_i{a^n_i\braket{(\Pi_{i})_w}_{\psi}^{b}},\label{26}
\end{align}
where `$n$' is any positive integer. Eqs. (\ref{25}) and (\ref{26}) can be solved to obtain the weak values of projection operators for different `$n$'. For example in three dimensional Hilbert Space, we require only up to second moment weak values because one can use the completeness relation for the projection operators with pre selection $\ket{\psi}$ and post-selection $\ket{b}$ 
 \begin{align}
1=\sum_i{\braket{(\Pi_{i})_w}_{\psi}^{b}}.\label{27}
\end{align}
In Appendix \ref{C}, we explicitly show  how to solve the above equations to obtain the weak values of the projection operators.\par
From (\ref{C1}), the highest moment weak value is $\braket{A^{d-1}_w}_{\psi}^{b}$ and as we have discussed in (\ref{II A}) that, extracting the highest moment weak value is enough to calculate all the lower moments weak values. \emph{Hence the total number of measurements needed to reconstruct a pure state is ${d}/{2}$ if the dimension $d$ is even and $({d-1})/{2}$ if the dimension $d$ is odd} (see \ref{II A} for detail discussion). \par
To compare with Lundeen \emph{et al.} \cite{29} and Wu \cite{47}, the number of measurement operators which is the complete set of projection operators  with a fixed post-selection is $(d-1)$. Moreover, we measure only one  system operator $A$, but post-selections are to be changed, while, in their method, there are $(d-1)$  system operators (projection operators) to be measured according to the AAV method.\par
Note that the weak values of projection operators in Eq. (\ref{22}) are exactly the weak-valued probabilities which were mentioned in the introduction section. \par
\normalsize{\textit{Alternative:–}}
The weak value of an observable $C$ with pre and post-selections $\ket{\psi}$ and $\ket{0}$, respectively  is
\begin{align}
\braket{C_w}_{\psi}^{0}=\frac{\braket{0|C|\psi}}{\braket{0|\psi}}.\label{28}
\end{align}
Inserting the identity operator $I=\sum_{i}{\ket{i}\bra{i}}$ in numerator of the right hand side of Eq. (\ref{28}), we have
\begin{align}
\braket{C_w}_{\psi}^{0} - C_{00}=\sum_{i=1}{C_{0i}\frac{\alpha_i}{\alpha_0}},\label{29}
\end{align}
where $C_{0i}=\braket{0|C|i}$. Like Eq. (\ref{29}), we have to measure a set of observables   to obtain the values of $\alpha_1/\alpha_0$,  $\alpha_2/\alpha_0$,$\cdots$,$\alpha_{(d-1)}/\alpha_0$  (see Appendix \ref{C}). The value of $\alpha_0$ can be obtained from the normalization condition. Here, the number of measurement operators is $(d-1)$. This method will be used to reconstruct an unknown quantum pure state of a bipartite system.
\subsubsection{Mixed state}
Measurement of a mixed state of a quantum system was also  introduced by  Lundeen \emph{et al.} \cite{4} using product weak values of two non commuting projection operators. It was later simplified by Shengjun Wu \cite{47} where weak values of complete set of  projection operators with complete set of post-selected states have been used.  Here, we develop another method by using the weak values of arbitrary observables which can be thought as the generalization of the reference  \cite{47}. This part is added here because we will use the same procedure while dealing with bipartite mixed state reconstruction using product weak values.\par
Let the unknown mixed state be the pre-selected state  then the weak value of the observable $C$ with post-selection $\ket{j}$ is 
\begin{align}
\braket{C_w}_{\rho}^{j}=\frac{\braket{j|C\rho|j}}{\braket{j|\rho|j}}.\label{30}
\end{align}
Inserting the identity operator $I=\sum_{i}{\ket{i}\bra{i}}$, we have
\begin{align}
p(\rho,j)\braket{C_w}_{\rho}^{j}=\sum_i{C_{ji}\rho_{ij}},\label{31}
\end{align}
where $p(\rho,j)=\braket{j|\rho|j}$ is the probability of obtaining the basis state $\ket{j}$ as a post-selection and $C_{ji}=\braket{j|C|i}$, $\rho_{ij}=\braket{i|\rho|j}$. To obtain the $j$-th column  of the density matrix $\rho$ from Eq. (\ref{31}), we need to measure a set of  arbitrary observables according to the AAV method to get a set of equations like  (\ref{31}) (see Appendix \ref{D}). \par
To compare with the work by Lundeen \emph{et al.} \cite{4} where each matrix element is directly obtainable via sequential measurements of two non-commuting projection operators in the AAV method. Different combinations of  position and momentum correlations are to be measured  where the correlations are of second order in terms  of the interaction coefficient. Our method is more efficient as we  only need to measure $(d-1)$ arbitrary  single observables  according to AAV method. We do not discard any post-measurements data and thus reduces the number of experimental runs (see Appendix \ref{D}). \par
Weak measurement methods have  several key advantages for state reconstruction of a quantum system  over the  standard schemes \cite{4}. For example, the state disturbance is minimum and thus it is possible for characterization of the state of a system during a physical  process in an experiment. Unlike standard schemes, global reconstruction is not required by our methods as states can be determined locally i.e., each matrix element can directly be accessed. Standard scheme typically requires $\mathcal{O}(d^2)$ measurements, while our method require  $\mathcal{O}(d-1)$ measurements for mixed state reconstruction.\par
Recently Vallone \emph{et al.} \cite{48} have shown: ``Strong Measurements Give a Better Direct Measurement of the Quantum Wave Function".  Namely, they have considered the von Neumannn Hamiltonian with basis projection operator (of the system) and Pauli operators (of the two dimensional pointer) and coupling coefficient (without approximation).  After the evolution of the system and the pointer,  the system is projected in the uniform superposition of the basis states. After that, each wavefunctions or basis coefficients of the concerned system state are calculated using the experimental probabilities obtained from the measurement of the  two dimensional pointer's different observables.    To compare,\\
{(i) In our method, the state disturbance is minimum. While in  the work of Vallone \emph{et al.} \cite{48}, the system will be disturbed strongly.} \\
(ii)  In the  method of Vallone \emph{et al.} \cite{48},   the post-selection of the system has to be of particular forms  to make the scheme successful otherwise (a)  the systematic error (trace distance) will be independent of the interaction coefficient which is one of their main concerns in the scheme (b)  wavefunctions for each computational basis will diverge and hence the scheme will fail.  Dimension of the pointer's Hilbert space is considered to  be two dimensional (finite dimensional). By such restrictions, the method  can only be used for limited number of quantum systems  (e.g., optical systems). While in our methods, there are no such restrictions on pointer states. The most suitable ways can be applied to obtain single weak values. \\
(iii) (a) In the  method of Vallone  \emph{et al.} \cite{48}, for pure state reconstruction of a single system, there are effectively `$d-1$' number of projection operators and for each projection operator, three different measurement observables are needed. So, the total number of measurement settings is `$3(d-1)$'. While in our method, the measurement settings are nearly `$d/2$' using higher moment weak values. (b) For the mixed state reconstruction, they need two independent pointers, three different joint pointer operators  and `$d-1$' number of projection operators and one post-selection in the system \cite{49}. Using such combinations, one need to calculate  mean values of  different combination of tripartite observables. In our method, `$d-1$' number of weak values  are required and there are no  such joint operations. \\
(iv)  Vallone  \emph{et al.} \cite{48}  have shown  that strong measurements outperform weak measurements  in both the ``precision and accuracy" for arbitrary quantum states in most cases. In our case, by performing the experiment many times on identically prepared systems, it is possible to reduce the uncertainty in the mean pointer displacement to any arbitrary precision \cite{50} (in order to obtain the  real and imaginary part of the weak values)\\
(v) For the given finite ensemble  size, our scheme can't give better performance than the methods of  \cite{48,49}  in terms of precision and accuracy \cite{51,52}. 
\subsection{State reconstruction of a bipartite system}
\subsubsection{Pure state}
In the above,  we introduced reconstruction of an unknown pure state of a given  system using single observable (or projection) weak values. For the  reconstruction of a bipartite  state, one needs to measure product  weak values namely the weak values of the tensor product observables.  In standard scheme i.e., von Neumann measurement scheme, measurement of product observables can not be realized directly  as it requires the interaction Hamiltonian of the two distant subsystems to be of the form $H\propto (A\otimes B)$  which implies an instantaneous interaction between the two distant subsystems (a relativistic constraint). In this section, we use our version of  product weak values (\ref{9}) in a bipartite system to  reconstruct  a    pure  state  following the same method of Eq. (\ref{29}) as we saw for the   pure state case in a single quantum system.\par
The pure state of a bipartite system can be written in computational basis as
\begin{align}
\ket{\psi_{AB}}=\sum_{ij}{\alpha_{ij}\ket{i_A}\otimes\ket{j_B}},\label{32}
\end{align}
where $\alpha_{ij}=\braket{ij|\psi_{AB}}$ and $\ket{ij}=\ket{i_A}\otimes\ket{j_B}$. 
 The product  weak value of the  observable  $C_{A}\otimes C_{B}$ in a bipartite system  is given by 
\begin{align}
  \braket{(C_{A}\otimes C_{B})_w}^{00}_{\psi_{AB}}=\frac{\braket{00|(C_{A}\otimes C_{B})|\psi_{AB}}}{\braket{00|\psi_{AB}}},\label{33}
\end{align}
where $\ket{\psi_{AB}}$ is the bipartite pre selected state which is to be reconstructed. $\ket{00}=\ket{0_A}\otimes\ket{0_B}$ is the post-selected state. Now inserting identity operator $I=\sum_{ij}{\ket{ij}\bra{ij}}$ of the joint Hilbert space in Eq. (\ref{33}),  we have
\begin{align}
\braket{( \hspace{-1pt}C_{A}\!\otimes \!C_{B} \hspace{-1pt})_w}^{00}_{\psi_{AB}} \hspace{-4pt}-  \hspace{-2pt}\big[C_{A}\big]_{00} \hspace{-1pt}\big[C_{B}\big]_{00} \hspace{-2pt}= \hspace{-8pt}\sum_{i,j\neq (0,0)} \hspace{-10pt}\big[C_{A}\big]_{0i} \hspace{-1pt}\big[C_{B}\big]_{0j}\frac{\alpha_{ij}}{\alpha_{00}}, \label{34}
\end{align}
 where  $\left[C_{A}\right]_{0i}\left[C_{B}\right]_{0j}=\braket{00|C_{A}\otimes C_{B}|ij}$ and $\alpha_{00}\neq 0$. Again we have to solve a matrix equation using  Eq. (\ref{34}) for a set of product operators  to obtain the values $\frac{\alpha_{ij}}{\alpha_{00}}$ (see Appendix \ref{E}) .\par
We have found  using the matrix equation (\ref{E1}) that we do not require to measure all the product weak values. For example,  consider a two-qubit system where 
\begin{equation}
\begin{split}
C^{(1)}_{A}\otimes C^{(1)}_{B}&=I^A\otimes \sigma_x^B ,\hspace{2mm}  C^{(2)}_{A}\otimes C^{(2)}_{B}=\sigma_x^A\otimes I^B\label{35}\\[10pt]
&C^{(3)}_{A}\otimes C^{(3)}_{B}=\sigma_x^A\otimes \sigma_x^B,
\end{split}
\end{equation}
then the square matrix in  (\ref{E1}) becomes an identity matrix having nonzero determinant and hence all the coefficients can be determined. \emph{This way, the number of measurements of product weak values can be reduced}. In this particular case, we need only one product weak value i.e., $\braket{(\sigma_x^A\otimes \sigma_x^B)_w}^{00}_{\psi_{AB}}$ to be measured and according to Eq. (\ref{9}), it can be calculated using local weak values $\braket{ (\sigma_x^B)_w}^{00}_{\psi_{AB}}$, $\braket{(\sigma_x^B)_w}^{10}_{\psi_{AB}}$ and $\braket{(\sigma_x^A)_w}^{00}_{\psi_{AB}}$.   So the total number of local weak values  is only three    to reconstruct two-qubit  pure state.  Note that, the local weak value  $\braket{(\sigma_x^B)_w}^{10}_{\psi_{AB}}$ can be calculated by using  $\braket{(\sigma_x^B)_w}^{00}_{\psi_{AB}}$ with the completeness relation for $\ket{0}$ and $\ket{1}$.\par
To compare with the method of Pan \emph{et al.} \cite{5}, our method  is experimentally simple because it  does not depend on the nature of  the pointer's state (entangled or product) and  locally measurable (using local weak values only) while in their method, the use of  entangled pointer's states are necessary (which might not be an easy task to perform) and local modular values as well as  modular values of the sum of the local operators are required. For certain cases, some of the probability amplitudes with the entangled pointer's states are considered to be sufficiently small.  Complicated situations may arise for higher dimensions and multi-partite systems because of the entangled pointer states. Our method can be generalized  both in higher dimensions and multi-partite systems with local weak value measurements only. In the method of  Pan \emph{et al.} \cite{5},  The number of product weak values is $(m-1)(n-1)$ and each product weak value  consists one  modular value of the sum of the two local projectors as well as two local projector modular values. Here `$m$' and `$n$' are the number of dimensions of the subsystems A and B, respectively. \emph{In our  method, there are  $(d-1)$ (where $d=mn$) numbers of product weak values and each product weak value can be extracted with only  two numbers of AAV type weak measurements}. But as we have seen for the case of two-qubit system (\ref{35}), we do not need to calculate  $(d-1)$ numbers of product weak values all the time. For example, effectively we need only two local weak values to reconstruct the pure state of the two-qubit system. Note that, in Ref. \cite{5} for two-qubit system, the total number of measurements is three in which one is the  modular value of the sum of two local projectors and two local projector modular values. So, in most of the cases, it is  possible to reduce  the  number of  product weak values  considerably  in our method of  state reconstruction.
\subsubsection{Mixed state}{}
To reconstruct a mixed state of a bipartite system, we will use the method of Eq. (\ref{31}).  Now, let the pre-selection of the system be $\rho_{AB}$ which is unknown and post-selection be any computational  basis state $\ket{kl}$. Then the product weak value of the   operator $C_A\otimes C_B$ is given by
\begin{align}
  \braket{(C_{A}\otimes C_{B})_w}^{kl}_{\rho_{AB}}=\frac{\braket{kl|\left(C_{A}\otimes C_{B}\right)\rho_{AB}|kl}}{\braket{kl|\rho_{AB}|kl}}.\label{36}
\end{align}
Now inserting the identity operator $I=\sum_{i,j}{\ket{ij}\bra{ij}}$ in Eq. (\ref{36}), we have
\begin{align}
  p(\rho_{AB},kl) \hspace{-2pt}\braket{( \hspace{-1pt}C_{A}\!\otimes\! C_{B} \hspace{-1pt})_w}^{kl}_{\rho_{AB}} \hspace{-2pt}= \hspace{-2pt}\sum_{i,j}\left[C_{A}\right]_{ki} \hspace{-1pt}\left[C_{B}\right]_{lj} \hspace{-1pt}[\rho_{AB}]_{ij,kl},\label{37}
\end{align}
where $\left[C_{A}\right]_{ki}\left[C_{B}\right]_{lj}=\braket{kl|C_{A}\otimes C_{B}|ij}$, $[\rho_{AB}]_{ij,kl}=\braket{ij|\rho_{AB}|kl}$ and $p(\rho_{AB},kl)=\braket{kl|\rho_{AB}|kl}$ is the probability of the successful post-selection $\ket{kl}$.
So to obtain the \emph{kl}-th column of the density matrix $\rho_{AB}$, we have to form a matrix equation using  Eq. (\ref{37}) for a set of product operators (see Appendix \ref{F}).\par
Clearly, mixed state reconstruction is more resource intensive than the pure state case in a bipartite system.  The number of product weak values to be calculated here is ($d-1$) and each product weak value can be extracted with only  three numbers of AAV type weak  measurements (see \ref{II B}). Here $d=m n$, where `m' and `n' are dimensions of the subsystems A and B, respectively.  \emph{We will  get advantage of using  matrix equation} (\ref{ F1}) \emph{where for some cases we do not require to calculate all the product weak values as we have seen  for the case of bipartite pure state reconstruction}. \par
It is  important to note  that, our method of calculating product weak values for pure and mixed states in a bipartite system can also be applied for  projection operators and hence one can reconstruct  pure  state using the state reconstruction method of Ref. \cite{5}. For mixed state reconstruction, one should look to the  Ref. \cite{47} by considering   bipartite system conditions.\par
Full knowledge of the state of a quantum system is always crucial  to understand a system better and for controlling quantum technologies. In particular, the measurement of bipartite (multi-partite) states are useful for information transfer, cryptography protocols, etc. They are also used to study nonlocality, quantum discord, entanglement entropy, etc.  We have shown the application of product and higher moment weak values as quantum state reconstruction of a single and bipartite systems only. The calculations of product weak values  of a bipartite system  are  even more  fascinating because of their local realizations. We can have  applications  of  product weak values  to extract informations about multi-partite systems for future technologies. Product weak values with local realization can find it's applications in quantum steering, to perform some nonlocal tasks, etc.
\section{Entanglement detection}\label{e IV}
Due to an immense application of  entangled systems \cite{53,54}, it is, by default,   an  important task in the field of quantum information to detect whether the shared states are entangled or not.  Here we show that product weak values (introduced in sec. \ref{e II}) can be used to detect entanglement of a bipartite system's state. Product weak values are experimentally accessible quantities and we have discussed in sec. \ref{e II} how one can do  that.    We have found  a necessary separability criteria for finite-dimensional systems.  By  clever choices of product observable and post-selections, it is possible to achieve the PPT criteria for entanglement detection of several important class of entangled bipartite states. Our method of entanglement detection can definitely be used for more class of entangled states.\par
 There are some existing  necessary separability criteria \cite{53,54} for detection of entanglement for finite-dimensional systems  based on local uncertainty relation (standard deviation based) \cite{55}, entropic uncertainty relations \cite{56}, separability inequalities on Bell correlations \cite{57} (which are exponentially stronger than the corresponding  local reality inequalities), etc. It is worth mentioning here that  Uffink and Seevink  provided  a single separability inequality \cite{58}, (although the choice of the observables being state-dependent)  quadratic in nature  is used to detect separability / entanglement  of all two-qubit states. \par
The separable states are considered to be of the following form 
\begin{align}
\rho=\sum_i{p_i\rho_A^i\otimes \rho_B^i}\label{38},
\end{align}
where $\rho_A^i=\ket{\psi_A^i}\bra{\psi_A^i}$, $\rho_B^i=\ket{\psi_B^i}\bra{\psi_B^i}$ and $\sum_i p_i=1$.  We will consider the following quantity which is directly connected to the product weak value (\ref{12}) for mixed states
\begin{widetext}
\begin{align}
&\left| \braket{\phi_A\phi_B|(A\otimes B)\rho|\phi_A\phi_B} \right|^2\nonumber\\
&=\left| \sum_i p_i\braket{\phi_A|A\rho_A^i|\phi_A} \braket{\phi_B|B\rho_B^i|\phi_B} \right|^2\nonumber\\
&=\left| \sum_i \left\{\sqrt{p_i}\frac{\braket{\phi_A|A\rho_A^i|\phi_A}}{\sqrt{\braket{\phi_A|\rho_A^i|\phi_A}}}\sqrt{\braket{\phi_B|\rho_B^i|\phi_B}}\right\} \left\{\sqrt{p_i}{\sqrt{\braket{\phi_A|\rho_A^i|\phi_A}}} \frac{\braket{\phi_B|B\rho_B^i|\phi_B}}{\sqrt{\braket{\phi_B|\rho_B^i|\phi_B}}}\right\}\right|^2\nonumber\\
& \leq \left( \sum_i p_i\frac{\left|\braket{\phi_A|A\rho_A^i|\phi_A}\right|^2}{\braket{\phi_A|\rho_A^i|\phi_A}}\braket{\phi_B|\rho_B^i|\phi_B} \right)\left( \sum_i p_i\braket{\phi_A|\rho_A^i|\phi_A}\frac{\left|\braket{\phi_B|B\rho_B^i|\phi_B}\right|^2}{\braket{\phi_B|\rho_B^i|\phi_B}} \right)\nonumber\\
&=\left( \sum_i p_i\braket{\phi_A|A|\psi_A^i}{\braket{\psi_A^i|A|\phi_A}}\braket{\phi_B|\psi_B^i}\braket{\psi_B^i|\phi_B}\right)\left( \sum_i p_i\braket{\phi_A|\psi_A^i}{\braket{\psi_A^i|\phi_A}}\braket{\phi_B|B|\psi_B^i}\braket{\psi_B^i|B|\phi_B}\right)\nonumber\\
&=\braket{\phi_A\phi_B|(A\otimes I)\rho(A\otimes I)|\phi_A\phi_B}\braket{\phi_A\phi_B|(I\otimes B)\rho(I\otimes B)|\phi_A\phi_B}\nonumber\\
&=\braket{\phi_A|A\rho_A^{\phi_B}A|\phi_A}\braket{\phi_B|B\rho_B^{\phi_A}B|\phi_B}\nonumber\\
&=\braket{\phi_A|A^2|\phi_A}\braket{\phi_B|B^2|\phi_B}\braket{\phi_A^{\prime}|\rho_A^{\phi_B}|\phi_A^{\prime}}\braket{\phi_B^{\prime}|\rho_B^{\phi_A}|\phi_B^{\prime}}\label{39},
\end{align}
where we have applied the Cauchy-schwarz inequality,  $\rho_X^{\phi_Y}=\braket{\phi_Y|\rho|\phi_Y}$ and $\ket{\phi_X^{\prime}}=X\ket{\phi_X}/\sqrt{\braket{\phi_X|X^2|\phi_X}}$ with $X\neq Y$ and X,Y=\{A,B\}.  The quantity $\braket{\phi_X^{\prime}|\rho_X^{\phi_Y}|\phi_X^{\prime}}$ can experimentally be obtained  in the following way.  At the first stage, measure the projection operator $\Pi_{\phi_Y}=\ket{\phi_Y}\bra{\phi_Y}$ in the subsystem `Y' on the shared bipartite state $\rho$. The collapsed state in the subsystem `X' is now $\rho_X^{\phi_Y}$, which is also the prepared state in this subsystem. At the second stage, measure the projection operator $\Pi_{\phi_X^{\prime}}=\ket{\phi_X^{\prime}}\bra{\phi_X^{\prime}}$  in the subsystem `X'. Note that, $\braket{\phi_X|X^2|\phi_X}$ can be obtained by just knowing the matrix form of the operator $X$ and $\ket{\phi_X}$.
\end{widetext}\par
The violation of the above inequality will imply entanglement of the given bipartite state.  Now, the following  examples will show the potential of the above inequality to detect entanglement of certain class of entangled states. \\
\textit{(i)} \textit{Two-qubit Werner state} (noisy singlet):
\begin{align}
\rho=p\ket{\psi_{AB}^-}\bra{\psi_{AB}^-}+(1-p)\frac{I_A\otimes I_B}{4},\nonumber
\end{align}
where $\ket{\psi_{AB}^-}=\frac{1}{\sqrt{2}}(\ket{01}-\ket{10})$. By choosing $A=\sigma^x_A$, $B=\sigma^x_B$, $\ket{\phi_A}=\ket{1}$ and $\ket{\phi_B}=\ket{0}$, it can be shown that the inequality is violated for $p>1/3$ (PPT criterion).\\
\textit{(ii)} \textit{Mixture of two Bell states}:
\begin{align}
\rho=p\ket{\phi_{AB}^+}\bra{\phi_{AB}^+}+(1-p)\ket{\phi_{AB}^-}\bra{\phi_{AB}^-},\nonumber
\end{align}
where $\ket{\phi_{AB}^+}=\frac{1}{\sqrt{2}}(\ket{00}+\ket{11})$ and $\ket{\phi_{AB}^-}=\frac{1}{\sqrt{2}}(\ket{00}-\ket{11})$.\par
Consider  $A=\sigma^x_A$, $B=\sigma^x_B$, $\ket{\phi_A}=\ket{1}$ and $\ket{\phi_B}=\ket{1}$. The inequality is violated for $p\neq 1/2$ (PPT criterion).\\
\emph{(iii)} The following density operator 
\begin{align}
\rho=p\ket{\psi_{AB}}\bra{\psi_{AB}} +(1-p)\frac{I_A\otimes I_B}{4}\nonumber
\end{align}  
where $\ket{\psi_{AB}}=a\ket{00}+b\ket{11}$ and $|a|^2+|b|^2=1$, is entangled if and only if $p > 1/(1+4|ab|)$ (PPT criterion).\par
Using the above separability criterion with the choices $A=\sigma_A^x$, $B=\sigma_B^x$,  $\ket{\phi_A}=\ket{1}$ and $\ket{\phi_B}=\ket{1}$. The inequality is violated for $p > 1/(1+4|ab|)$.\\
\textit{(iv)} The density operator 
\begin{align}
\rho=p\ket{\psi_{AB}^{(1)}}\bra{\psi_{AB}^{(1)}} +(1-p)\ket{\psi_{AB}^{(2)}}\bra{\psi_{AB}^{(2)}}\nonumber
\end{align}  
where $\ket{\psi_{AB}^{(1)}}=b_1\ket{01}+c_1\ket{10}$, $\ket{\psi_{AB}^{(2)}}=b_2\ket{01}+c_2\ket{10}$,  $|b_1|^2+|c_1|^2=1$ and $|b_2|^2+|c_2|^2=1$, is entangled if and only if (PPT criterion)  $|pb_1^*c_1+(1-p)b_2^*c_2| > 0$.\par 
Using the  separability criterion with the choices $A=\sigma_A^x$, $B=\sigma_B^x$,  $\ket{\phi_A}=\ket{0}$ and $\ket{\phi_B}=\ket{1}$. The inequality is violated for $|pb_1^*c_1+(1-p)b_2^*c_2| > 0$.\\
\textit{(v) Mixture of 4-Bell states:} 
\begin{align}
\rho=&p_1\ket{\psi_{AB}^+}\bra{\psi_{AB}^+}+p_2\ket{\psi_{AB}^-}\bra{\psi_{AB}^-}\nonumber\\&
+p_3\ket{\phi_{AB}^+}\bra{\phi_{AB}^+}+p_4\ket{\phi_{AB}^-}\bra{\phi_{AB}^-},\nonumber
\end{align}
where $\ket{\psi_{AB}^+}=\frac{1}{\sqrt{2}}(\ket{00}+\ket{11})$,  $\ket{\psi_{AB}^-}=\frac{1}{\sqrt{2}}(\ket{00}-\ket{11})$ and $p_1+p_2+p_3+p_4=1$. This density matrix is entangled if and only if $p_i >1/2$, $p_j < 1/2$, $i\neq j$ and $i,j=1,2,3,4$ (PPT criterion).\par
Consider (a) $A=\sigma^x_A$, $B=\sigma^x_B$, $\ket{\phi_A}=\ket{0}$ and $\ket{\phi_B}=\ket{0}$. The inequality is violated for $p_1> 1/2$ or $p_2>1/2$, (b) $A=\sigma^x_A$, $B=\sigma^x_B$, $\ket{\phi_A}=\ket{0}$ and $\ket{\phi_B}=\ket{1}$. The  inequality is violated for $p_3> 1/2$ or $p_4>1/2$.\\
\textit{(vii) Two qudit Werner  states \cite{59}:}
\begin{align}
\rho=(1-p)\frac{2}{d^2+d}P^{(+)} + p\frac{2}{d^2-d}P^{(-)}, \hspace{3mm} 0\leq p\leq 1,\nonumber
\end{align}
where the projectors $P^{(+)}=(I+V)/2$, $P^{(-)}=(I-V)/2$ with identity $I$ and flip operation $V=\sum_{i,j=0}^{d-1}{\ket{i}\bra{j}\otimes \ket{j}\bra{i}}$, and $\{\ket{i}\}$ is the basis states. The state $\rho$  is entangled if and only if  $p>1/2$  (PPT criterion).\par
To see, which values of `$p$' are achievable via the separability inequality Eq. (\ref{39}), we first calculate the entanglement condition and then will see some physically implementable systems. Consider $\bra{i^{\prime}_A i^{\prime}_B}C_A\otimes C_B=\bra{j^{\prime}_A j^{\prime}_B}$, where  $\braket{i^{\prime}_A|j^{\prime}_A}=0$ or $\braket{i^{\prime}_B|j^{\prime}_B}=0$ or both. Then from Eq. (\ref{39}), the LHS - RHS   becomes
\begin{align}
|\Lambda^{(-)}\braket{j^{\prime}_A|i^{\prime}_B}\braket{j^{\prime}_B|i^{\prime}_A}\hspace{-1mm}|^2 - &
\left[\Lambda^{(+)}+\Lambda^{(-)}|\hspace{-1mm}\braket{j^{\prime}_A|i^{\prime}_B}\hspace{-1mm}|^2\right]\nonumber\\
& \times\left[\Lambda^{(+)}+\Lambda^{(-)}|\hspace{-1mm}\braket{j^{\prime}_B|i^{\prime}_A}\hspace{-1mm}|^2\right]\label{40}
\end{align}
where $\Lambda^{(+)}=\frac{1-p}{d^2+d} + \frac{p}{d^2-d}$ and $\Lambda^{(-)}=\frac{1-p}{d^2+d} - \frac{p}{d^2-d}$. Now by making the choices $|\hspace{-1mm}\braket{j^{\prime}_B|i^{\prime}_A}\hspace{-1mm}|=|\hspace{-1mm}\braket{j^{\prime}_A|i^{\prime}_B}\hspace{-1mm}|=1$, it is easy to show that LHS-RHS (\ref{40}) is alway positive for $p> \frac{3(d-1)}{2(2d-1)}$ which is the entanglement condition for the Werner state in $d\otimes d$. It known that for $\frac{1}{2}<p\leq \frac{3(d-1)}{2(2d-1)}$ and $p> \frac{3(d-1)}{2(2d-1)}$, the Werner state is bound entangled (conjectured) and distillable respectively.  That is, our separability criterion (\ref{39}) is able to detect the distillability   of the Werner state but not the bound entanglement (if any). In Appendix \ref{G}, we give the examples of how to fulfil  the choices we made here in the physical systems.\\
\textit{(vi) Higher dimensional isotropic states \cite{60}:}
\begin{align}
\rho=p\ket{\psi^+_{AB}}\bra{\psi^+_{AB}}+(1-p)\frac{I_A\otimes I_B}{d^2},\nonumber
\end{align}
where $\ket{\psi_{AB}^+}=\frac{1}{\sqrt{d}}(\sum_{i=1}^d\ket{i_Ai_B}$ and `$d$' is the dimension of the subsytems. \par 
By choosing the spin flip operators  $A=\sigma^x_A$, $B=\sigma^x_B$ such that $(\sigma^x_A\otimes \sigma_B^x)\ket{i_Ai_B}=\ket{j_Aj_B}$, $j_A\neq i_A$, $j_B\neq i_B$,  and $\ket{\phi_A}=\ket{i_A}$,  $\ket{\phi_B}=\ket{i_B}$, it can be shown that the inequality is violated for $p>1/(d+1)$ (PPT criterion).\par
In comparison with most of the existing works, the above separability inequality is easier to implement in experiments due to the simple realization of weak measurement and less number of measurement settings. In particular, compared to the case of universal (i.e., state-independent) detection of two-qubit entanglement using two copies of the state at a time and using the notion of weak values \cite{37}, the aforesaid inequality (\ref{39}) (involving product weak values) uses only a single copy of the bipartite state ${\rho}$ at a time. Moreover, the criterion is resource-wise better than tomography, based on local realization and dependent on one type of measurement set-up. \par
We do believe that the separability inequality (\ref{39}) is of universal nature  at least for the set of all two-qubit states. Needless to say that the choice of the local observables $A$, $B$ as well as the post-selected state $\ket{{\phi}_A{\phi}_B}$ do depend upon the choice of the input bi-partite state ${\rho}$.
\section{Robustness}\label{e V}
In AAV method, the coupling between  the system and the pointer is extremely small and hence the state collapse  is avoided. During the process, any resolution is insufficient to distinguish the different eigenvalues of the observable. Nevertheless, by performing the experiment many times on identically prepared systems, it is possible to  reduce the uncertainty in the mean pointer displacement to any arbitrary precision \cite{50}. \par 
There  are other type of errors which  are inevitable due to   the inappropriate  choices of system observables and unsharp post-selections. Here, we show that our methods of  ``extraction of product and higher moment weak values" are robust against them.\\ 
\textsl{(i)} \normalsize{\textit{Error in choice of observable:}}  In experiment, let's say, we want to measure a spin-1/2 observable according to the AAV method but due to some  technical difficulties, we are  unable to  measure the actual spin-1/2 observable (slightly changed $\theta$ and $\phi$, where $\theta$ and $\phi$ define a point on the bloch sphere). Now let `$A$' be the correct observable while `$A^e$' be the erroneous one such that  $\left|A-A^e\right| \leq \delta$, where $\left|X\right| = Tr\sqrt{X^{\dagger}X}$   is the trace norm of a square matrix X. So the error occurring  in the weak value is given by 
\begin{align}
\Delta(\rho,A,\phi)&=\left|{\braket{A_w}^{\phi}_{\rho} - \braket{A^e_w}^{\phi}_{\rho}}\right|=\frac{\left|\braket{\phi|(A-A^e)\rho|\phi}\right|}{\braket{\phi|\rho|\phi}}\nonumber\\
&\leq \frac{\left|\braket{\phi|(A-A^e)\rho|\phi}\right|}{m}, \label{41}
\end{align}
where `m' is the minimum of the probabilities for all the possible choices  of rank-one post-selections with a given pre-selection. Now consider the spectral decomposition $A-A^e=\sum_i{\lambda_i\ket{i}\bra{i}}$ where $\{\ket{i}\}$ is the complete set of orthogonal basis. Then
\begin{align}
\Delta(\rho,A,\phi)&\leq\frac{1}{m}\left|\sum \lambda_i \braket{\phi|i}\braket{i|\rho|\phi} \right|\nonumber\\
&\leq\frac{1}{m}\sum_i \left|\lambda_i\right| \left| \braket{\phi|i}\braket{i|\rho|\phi}\right|. \label{42}
\end{align}
Note that $\left|A-A^e\right|=\sum_i \left|\lambda_i\right|$ and since $0\leq \rho \leq \mathbbm{1} $, it can easily be shown that $\left| \braket{\phi|i}\braket{i|\rho|\phi}\right|\leq 1$. Hence
\begin{align}
\Delta(\rho,A,\phi)&\leq\frac{1}{m}\sum_i \left|\lambda_i\right|=\frac{\left|A-A^e\right|}{m}\leq \frac{\delta}{m}. \label{43}
\end{align}
\textsl{(ii)} \normalsize{\textit{Noisy post-selection:}} Now we consider another type of error which is common  in experiments is due to the  unsharp post-selections. Let us assume that the unsharp post-selection is a mixture of the true post-selection $\ket{\phi}$ with probability $(1-\epsilon)$ and noise state $\sigma$ with probability $\epsilon$ 
\begin{align}
 \Phi^{\epsilon}=(1-\epsilon)\ket{\phi}\bra{\phi} + \epsilon\sigma,\label{44}
\end{align}
where $\epsilon$ is a sufficiently small positive quantity. Then the difference between the perturbed and true weak values is 
\begin{align}
\braket{A_w}^{ \Phi^{\epsilon}}_{\rho} - \braket{A_w}^{\phi}_{\rho}\approx\epsilon\left[ \frac{Tr(\sigma A \rho) - \braket{A_w}^{\phi}_{\rho} Tr(\sigma\rho)}{\braket{\phi|\rho|\phi}} \right].\label{45}
\end{align}\par
So in both the cases (Eqs. (\ref{43}) and (\ref{45})), the weak values are robust. Now it is not hard to realize that product and higher moment weak values are also robust. The only thing we need to do is to replace the observable $A$ by  $A^2$ for a single system and  $C_A\otimes C_B$ for a bipartite system in  Eqs. (\ref{43}) and (\ref{45}). Hence the weak values which we have used to reconstruct the state of a single and bipartite systems are also robust.\\
\section{Conclusion}\label{e VI}
We have derived the methods  of extracting  higher moment weak values and product weak values using Vaidman's formula.  Such higher moment weak values are  calculated using only the weak values of that observable with pairwise orthogonal post-selections.  Two dimensional Hilbert space becomes the simplest case for extracting the  higher moment weak values. Our methods turn out to be  simple from experimental perspective as we don't need to measure the N pointer's correlations as required in the previous works. Previously, it was thought that with  Gaussian pointers' states, it is not possible to obtain the higher moment weak values but we have shown that  instead of looking for different pointer states (e.g., OAM states)   to obtain the higher moment weak values, we  can use  Vaidman's formula. To extract the product weak values in a bipartite system, we have again used  Vaidman's formula in one of the subsystems. The product weak values can be calculated using only local weak values. The key factor for such local realization is that the action of the local operator on the local post-selected state is equivalent to the superposition of that post-selected state and a unique orthogonal state to that given post-selected state. Our method can be used to verify  Hardy’s Paradox, to confirm the existence of quantum Cheshire cats, to perform EPR-Bohm experiment, to realize non-locality via post-selections, etc.\par
As an application, we have shown how to  reconstruct quantum states of a single and bipartite systems separately. We have used higher moment weak values to reconstruct  an unknown pure  state of a single system. The number of measurements are nearly half of the measurements  required in previous works. Mixed state reconstruction has been shown using arbitrary observbles.  We have used product weak values for  reconstruction  of   pure and mixed  states in a bipartite system. Such reconstructions become simply feasible in experiment using only the measurements of local weak values.  In the previous works, projection measurement operators were the central for direct quantum state tomography. But we have generalized it to any arbitrary observables for both single and bipartite systems. Comparisons between the previous works and our work have been  considered from various perspective (e.g., number of measurements according to the  AAV method and experimental feasibility).  { A necessary separability criteria (in terms of an inequality) for finite dimensional bi-partite systems using  product weak values has been derived. This inequality is turned out to be strong as the PPT criteria can be achieved for  certain class of entangled states  by cleverly choosing the product observables and the post selections.   The criteria can, in principle detect more classes of entangled states with suitably choosing product observables and post-selections}.  Finally, we have shown that our methods are robust against   the errors which  are inevitable due to   the inappropriate  choices of system observables and unsharp  post-selections. Our method can easily be extended to multi-partite systems.\par
{\bf{Acknowledgment}}: Sahil is thankful to the QIC group at HRI, Allahabad, for making an arrangement for visiting the group during which, part of the work was done. SM acknowledges financial support from the Visiting Postdoctoral Programme of IMSc, Chennai. We would like to thank  AK Pan and PK Panigrahi for bringing an useful comment \cite{43} on higher moment weak values calculation to our attention. 
\bibliographystyle{apsrev4-1}
\bibliography{Reference}

\section{APPENDICES}
\appendix
\section{}\label{A}
Let $A$ be a Hermitian operator  and $\ket{\phi}$ be a quantum state vector in a Hilbert space $\mathcal{H}$. Then  it is always possible to decompose ${A}\ket{\phi}$ as \cite{40}
\begin{align}
{A}\ket{\phi}=\alpha\ket{\phi} + \beta\ket{\phi^{\perp}_{A}},\label{A1}
\end{align}
where $\alpha$ and $\beta$ can be any complex numbers. $\ket{\phi^{\perp}_{A}}$ is an orthonormal state to $\ket{\phi}$. The state  $\ket{\phi_{A}^\perp}$ is unique and can be determined using Gram-Schmidt orthonormalization process (GSOP) as given below.\par 
Let $A\ket{\phi}$ and $\ket{\phi}$  are two non-orthogonal state vectors in the same Hilbert space. Using GSOP we find the unnormalized $\ket{\phi^{\perp}_{A}}_{un}$
\begin{equation}
 \ket{\phi_{A}^{\perp}}_{un}= {A}\ket{\phi} - \frac{(\braket{\phi |{A})|\phi} }{\braket{\phi|\phi}}\ket{\phi}=({A}-\braket{A}_{\phi})\ket{\phi}.\label{A2}
\end{equation}
Here $\ket{\phi}$ is normalized. So the normalized orthogonal state vector to $\ket{\phi}$ is
\begin{equation}
\ket{\phi^{\perp}_{A}}=\frac{\ket{\phi_{A}^{\perp}}_{un}}{\sqrt{_{un}\braket{\phi_{A}^{\perp}|\phi_{A}^{\perp}}_{un}}}.\label{A3}
\end{equation}
From Eqs. (\ref{A2}) and (\ref{A3})  we find $\alpha=\braket{\phi |{A}|\phi}=\braket{{A}}_{\phi}$ and $\beta=\braket{\phi^{\perp}_{A} |{A}|\phi}=\braket{\Delta A}_{\phi}$. Where $\braket{{A}}_{\phi}$ and $\braket{\Delta A}_{\phi}$ are the average and standard deviation of the observable ${A}$ in the state $\ket{\phi}$, respectively.  So the Eq. (\ref{A1}) becomes 
\begin{align}
{A}\ket{\phi}=\braket{{A}}_{\phi}\ket{\phi} + \braket{\Delta A}_{\phi} \ket{\phi^{\perp}_{A}}.\label{A4}
\end{align}
Eq. (\ref{A4}) is sometimes referred as Vaidman's formula.
\section{}\label{B}
For a given mixed pre selected state $\rho$ and  post-selected state $\ket{\phi}$, the second moment weak value of the observable $A$ is given by 
\begin{align}
\braket{{({A^2})}_w}^{\phi}_{\rho}&={\left(\frac{\braket{\phi|\rho A^2|\phi}}{\braket{\phi|\rho|\phi}}\right)}^{\ast}\nonumber\\
&=\braket{A}_{\phi}\frac{\braket{\phi|A\rho|\phi}}{\braket{\phi|\rho|\phi}} + \braket{\Delta A}_{\phi}\frac{{\braket{\phi_A^{\perp}|A\rho|\phi}}}{\braket{\phi|\rho|\phi}}.\label{B1}
\end{align}
Now  ${\braket{\phi_A^{\perp}|A\rho|\phi}}=Tr\big(\ket{\phi}\bra{\phi^{\perp}_{A}}A\rho\big)$ can be calculated as
\begin{align}
{\braket{\phi_A^{\perp}|A\rho|\phi}} + {\braket{\phi|A\rho|\phi_A^{\perp}}}=Tr\left[\big(\hspace{-1pt}\ket{\phi}\bra{\phi^{\perp}_{A}} + \ket{\phi^{\perp}_A}\bra{\phi}\hspace{-1pt}\big)A\rho\right],\label{B2}
\end{align}
 where $\big(\ket{\phi}\bra{\phi^{\perp}_{A}} + \ket{\phi^{\perp}_A}\bra{\phi}\big)$ is a normal operator satisfying $XX^{\dagger}=X^{\dagger}X$ (where X is any operator) and hence can be written in spectral decomposition  
\begin{align}
\ket{\phi}\bra{\phi^{\perp}_{A}} + \ket{\phi^{\perp}_A}\bra{\phi}=\sum_i^{d}{\lambda_i \ket{i}\bra{i}},\label{B3}
\end{align}
where $\ket{i}$ is the eigenvector with eigenvalue $\lambda_i$ and `$d$' is the dimension of the Hilbert space. Using Eq. (\ref{B3}) in (\ref{B2}), we have
\begin{align}
{\braket{\phi_A^{\perp}|A\rho|\phi}} + {\braket{\phi|A\rho|\phi_A^{\perp}}}&=\sum_i^{d}{\lambda_i \braket{i|A\rho|i}}\nonumber\\
&=\sum_i^{d}{\lambda_i \braket{A_w}^{i}_{\rho}p(\rho,i)},\label{B4}
\end{align}
where $p(\rho,i)=\braket{i|\rho|i}$ is the probability of obtaining the eigenvector $\ket{i}$ and $\braket{A_w}^{i}_{\rho}$ is the weak value of the observable $A$ for the given pre and post-selections $\rho$ and $\ket{i}$, respectively. Now similarly,
\begin{align}
{\braket{\phi_A^{\perp}|A\rho|\phi}} - {\braket{\phi|A\rho|\phi_A^{\perp}}}&=\sum_i^{d}{\lambda^{\prime}_i \braket{A_w}^{i^{\prime}}_{\rho}p(\rho,i^{\prime})},\label{B5}
\end{align}
\begin{widetext}
where we have used the fact that $\ket{\phi}\bra{\phi^{\perp}_{A}} - \ket{\phi^{\perp}_A}\bra{\phi}$ is also a normal operator with the spectral decomposition
\begin{align}
\ket{\phi}\bra{\phi^{\perp}_{A}} - \ket{\phi^{\perp}_A}\bra{\phi}=\sum_i^{d}{\lambda^{\prime}_i \ket{i^{\prime}}\bra{i^{\prime}}}.\label{B6}
\end{align}
Now adding two equations (\ref{B4}) and (\ref{B5}), we have 
\begin{align}
{\braket{\phi_A^{\perp}|A\rho|\phi}}=\frac{1}{2} \sum_i^{d}\hspace{-1pt}\left({\lambda_i \braket{A_w}^{i}_{\rho}p(\rho,i) + \lambda^{\prime}_i \braket{A_w}^{{i^\prime}}_{\rho}p(\rho,i^{\prime})}\hspace{-1pt} \right).\label{B7}
\end{align}
So the final expression of Eq. (\ref{B1}) is
\begin{align}
\braket{{({A^2})}_w}^{\phi}_{\rho}&=\braket{A}_{\phi}\braket{A_w}^{\phi}_{\rho} + \frac{\braket{\Delta A}_{\phi}}{2p(\rho,\phi)}\sum_i^{d}\left({\lambda_i \braket{A_w}^{i}_{\rho}p(\rho,i) + \lambda^{\prime}_i \braket{A_w}^{i^{\prime}}_{\rho}p(\rho,i^{\prime})} \right),\label{B8}
\end{align}
where $p(\rho,\phi)=\braket{\phi|\rho|\phi}$. The $n$-th moment weak value is 
\begin{align}
\braket{{({A^n})}_w}^{\phi}_{\rho}&=\braket{A}_{\phi}\braket{A^{n-1}_w}^{\phi}_{\rho} + \frac{\braket{\Delta A}_{\phi}}{2p(\rho,\phi)}\sum_i^{d}\left({\lambda_i \braket{A^{n-1}_w}^{i}_{\rho}p(\rho,i) + \lambda^{\prime}_i \braket{A^{n-1}_w}^{i^{\prime}}_{\rho}p(\rho,i^{\prime})} \right).\label{B9}
\end{align}
\end{widetext}\par
 Note that, $\{\braket{A_w}^{\phi}_{\rho}\}^{d}_{i=1}$ can be measured in the same experimental setup according to the AAV method as the post-selections \{$\ket{i}$\} form a complete set of orthogonal basis. Similarly, $\{\braket{A_w}^{\phi}_{\rho}\}^{d}_{i^{\prime}=1}$ can also be measured within the same experiment with the complete set of post-selections \{$\ket{i^{\prime}}$\}  according to the AAV method. So, to extract the second moment weak value, one needs only three AAV type measurements and these are $\braket{A_w}^{\phi}_{\rho}$,  $\{\braket{A_w}^{\phi}_{\rho}\}^{d}_{i=1}$ and  $\{\braket{A_w}^{\phi}_{\rho}\}^{d}_{i^{\prime}=1}$. Number of measurements will increase for  mixed state case than when the system is prepared in the pure state.
\section{}\label{C}
For the  measurement of a  pure state of a single system, we have followed a method  slightly different  from Lundeen \emph{et al.} \cite{29}. Namely we consider higher moment weak values as discussed in the main text. Here we show how to complete the process. $A$ is an observable having spectral decomposition (\ref{24}), is measured according to AAV method. Now consider Eq. (\ref{25}) and (\ref{26}) with $n$=1,$\dots$$d-1$, where $d$ is the dimension of the Hilbert space. So, for different `$n$', we obtain a matrix equation
{\scriptsize{
\begin{equation} \label{C1}
\renewcommand\arraystretch{1.8}
    \begin{pmatrix}
    1\\
    \braket{A_w}_{\psi}^{b}\\
    \vdots\\
    \braket{A^{d-1}_w}_{\psi}^{b}\\
    \end{pmatrix}
    = 
    \begin{pmatrix}
    1&1&\cdots&1\\
    a_0&a_1& &a_{d-1}\\
     \vdots&&\ddots\\
    a^{d-1}_0&a^{d-1}_1&\cdots &a^{d-1}_{d-1}\\
    \end{pmatrix}
     \begin{pmatrix}
    \braket{(\Pi_{0})_w}_{\psi}^{b}\\
   \braket{(\Pi_{1})_w}_{\psi}^{b}\\
    \vdots\\
   \braket{(\Pi_{{d-1}})_w}_{\psi}^{b}\\
    \end{pmatrix}
\end{equation}  }}
The solution of the above equation exists if the determinant of the square matrix  is non-zero.   The weak values of the projection operators are then extracted from (\ref{C1}) using the higher moment weak values of $A$.\\
\normalsize{\textit{Alternative:–}} 
To obtain the values of ${\alpha_i}/{\alpha_0}$, we have to solve the matrix equation using the Eq. (\ref{29}) for a set of  arbitrary observables $C^{(n)}, n=1,\dots,(d-1)$. So
{\scriptsize{
\begin{equation} \label{C2}
 \renewcommand\arraystretch{1.8}
    \begin{pmatrix}
    \braket{C^{(1)}_w}_{\psi}^{0}-C^{(1)}_{00}\\
    \braket{C^{(2)}_w}_{\psi}^{0}-C^{(2)}_{00}\\
    \vdots\\
    \braket{C^{(d-1)}_w}_{\psi}^{0}-C^{(d-1)}_{00}\\
    \end{pmatrix}\hspace{-4pt}
    =\hspace{-4pt} \begin{pmatrix}
    C^{(1)}_{01}&C^{(1)}_{02}&\cdots&C^{(1)}_{0(d-1)}\\
    C^{(2)}_{01}&C^{(2)}_{02}& &C^{(2)}_{0(d-1)}\\
     \vdots&&\ddots\\
   C^{(d-1)}_{01}&C^{(d-1)}_{02}&\cdots &C^{(d)}_{0(d-1)}\\
    \end{pmatrix}
     \begin{pmatrix}
    \frac{\alpha_1}{\alpha_0}\\
   \frac{\alpha_2}{\alpha_0}\\
    \vdots\\
   \frac{\alpha_{d-1}}{\alpha_0}\\
    \end{pmatrix}
\end{equation}} }\\
The solution exists if the  determinant  of the square matrix above is non-zero. It may be noted here that there will always exist at least one set of observables $\{C^{(n)}\}^{d-1}_{n=1}$ for which the determinant of the coefficient matrix on the right hand side of (\ref{C2}) is non-zero. Now we obtain the values of $\alpha_i/\alpha_0$ and then from Eq. (\ref{21}), we have
\begin{align}
\ket{\psi}=\alpha_0 \left(\ket{0}+\sum_{i=1}\frac{{\alpha_i}}{\alpha_0}\ket{i}\right).\label{C3}
\end{align}
 The value of $\alpha_0$  can be obtained from the normalization condition. If the determinant of the square matrix (\ref{C2}) is zero  with the post-selection $\ket{0}$, then one should look for the other post-selections such as $\ket{1}, \ket{2}$ and so on and hence solve the corresponding matrix equations.
\section{}\label{D}
 The $j$-th column of the density matrix with post-selection $\ket{j}$ can be obtained using Eq. (\ref{31}) for a set of  arbitrary observables $C^{(n)}, n=1,\dots,d$  as 
 {\scriptsize{
\begin{equation} \label{D1}
\renewcommand\arraystretch{1.8}
p(\rho,j)
    \begin{pmatrix}
    \braket{C^{(1)}_w}_{\rho}^{j}\\
    \braket{C^{(2)}_w}_{\rho}^{j}\\
    \vdots\\
    \braket{C^{(d)}_w}_{\rho}^{j}\\
    \end{pmatrix}
    = \begin{pmatrix}
    C^{(1)}_{j0}& \hspace{2mm} C^{(1)}_{j1}&\cdots&C^{(1)}_{j(d-1)}\\
    C^{(2)}_{j0}& \hspace{2mm} C^{(2)}_{j1}& &C^{(2)}_{j(d-1)}\\
     \vdots&&\ddots\\
   C^{(d)}_{j0}& \hspace{2mm} C^{(d)}_{j1}&\cdots &C^{(d)}_{j(d-1)}\\
    \end{pmatrix}
     \begin{pmatrix}
   \rho_{0j}\\
   \rho_{1j}\\
    \vdots\\
  \rho_{(d-1)j}\\
    \end{pmatrix}
\end{equation}}}
With $C^{(1)}=I$, an identity operator. The $j'$-th column can be calculated measuring the same set of observables with post-selection $\ket{{j{'}}}$. Remember that post-selections are complete set of projection operators and hence can be measured simultaneously for each observables $C^{(n)}$. So measuring an observable $C^{(n)}$ according to AAV method, we obtain all the weak values for a complete set of post selections $\{\ket{j}\}$.  We do not need to perform any measurement  for the last column  as one can get it using normalization and Hermiticity condition of the density matrix.
\section{}\label{E}
In the main text, we  discussed about how to develop an equation (\ref{34}) which contains all the probability amplitudes of a  pure state in a bipartite system of dimension $d=mn$, where `$m$' and `$n$' are the dimensions of subsystems A and B, respectively. Now, our task is to obtain all the unknown complex coefficients (probability amplitudes) in a pure state. To do so, we develop a matrix equation using Eq. (\ref{34}) for a set of  arbitrary product observables $C^{(n)}_{A}\otimes C^{(n)}_{B}, n=1,\dots,(d-1)$
\begin{widetext}
\begin{equation} {\label{E1}}
\renewcommand\arraystretch{2.3}
\scriptsize{
    \begin{pmatrix}
   \braket{(C^{(1)}_{A}\otimes C^{(1)}_{B})_w}^{00}_{\psi_{AB}} - \big[C^{(1)}_{A}\big]_{00}\big[C^{(1)}_{B}\big]_{00}\\
    \braket{(C^{(2)}_{A}\otimes C^{(2)}_{B})_w}^{00}_{\psi_{AB}} - \big[C^{(2)}_{A}\big]_{00}\big[C^{(2)}_{B}\big]_{00}\\
     \vdots\\
    \braket{(\hspace{-1pt}C^{(\hspace{-1pt}d-1\hspace{-1pt})}_{A}\hspace{-3pt}\otimes \hspace{-2pt}C^{(\hspace{-1pt}d-1\hspace{-1pt})}_{B}\hspace{-1pt})_w}^{00}_{\psi_{AB}}\hspace{-5pt}-\hspace{-2pt}\big[\hspace{-1pt}C^{(\hspace{-1pt}d-1\hspace{-1pt})}_{A}\hspace{-1pt}\big]_{00}\hspace{-1pt}\big[\hspace{-1pt}C^{(\hspace{-1pt}d-1\hspace{-1pt})}_{B}\hspace{-1pt}\big]_{00}
    \end{pmatrix}
   \hspace{-3pt} =\hspace{-3pt}
    \begin{pmatrix}
    \big[C^{(1)}_{A}\big]_{00}\big[C^{(1)}_{B}\big]_{01}&  \big[C^{(1)}_{A}\big]_{01}\big[C^{(1)}_{B}\big]_{00}&\cdots&
    \big[C^{(1)}_{A}\big]_{0(d-1)}\big[C^{(1)}_{B}\big]_{0(d-1)}\\
      \big[C^{(2)}_{A}\big]_{00}\big[C^{(2)}_{B}\big]_{01}& \big[C^{(2)}_{A}\big]_{01}\big[C^{(2)}_{B}\big]_{00}&&
      \big[C^{(2)}_{A}\big]_{0(d-1)}\big[C^{(2)}_{B}\big]_{0(d-1)}\\
       \vdots&&\ddots\\
        \big[\hspace{-1pt}C^{(\hspace{-1pt}d-1\hspace{-1pt})}_{A}\hspace{-1pt}\big]_{00}\hspace{-1pt}\big[\hspace{-1pt}C^{(\hspace{-1pt}d-1\hspace{-1pt})}_{B}\hspace{-1pt}\big]_{01}& \hspace{-3pt} \big[\hspace{-1pt}C^{(\hspace{-1pt}d-1\hspace{-1pt})}_{A}\hspace{-1pt}\big]_{01}\big[\hspace{-1pt}C^{(\hspace{-1pt}d-1\hspace{-1pt})}_{B}\hspace{-1pt}\big]_{00}&\hspace{-3pt}\cdots&\hspace{-1pt} \big[\hspace{-1pt}C^{(\hspace{-1pt}d-1\hspace{-1pt})}_{A}\hspace{-1pt}\big]_{0(\hspace{-1pt}d-1\hspace{-1pt})}\big[\hspace{-1pt}C^{(\hspace{-1pt}d-1\hspace{-1pt})}_{B}\hspace{-1pt}\big]_{0(\hspace{-1pt}d-1\hspace{-1pt})}
         \end{pmatrix}
     \begin{pmatrix}
    \frac{\alpha_{01}}{\alpha_{00}}\\
    \frac{\alpha_{10}}{\alpha_{00}}\\
     \vdots\\
     \frac{\alpha_{(\hspace{-1pt}d-1\hspace{-1pt})\hspace{-1pt}(\hspace{-1pt}d-1\hspace{-1pt})}}{\alpha_{00}}
        \end{pmatrix}}
       \end{equation}  
\end{widetext}
Solution exists if the determinant of the square matrix in (\ref{E1}) is nonzero and hence we calculate all the unknown coefficients $\alpha_{ij}/\alpha_{00}$. Then Eq. (\ref{31}) becomes
\begin{align}
\ket{\psi_{AB}}=\alpha_{00} \left(\ket{00}+\sum_{i,j\neq (0,0)}\frac{{\alpha_{ij}}}{\alpha_{00}}\ket{ij}\right).\label{E2}
\end{align}
Again, the value of $\alpha_{00}$ can be found using normalization condition. If the determinant of the square matrix becomes zero with the post selection $\ket{00}$ then one should look for the other post-selections such as $\ket{01}$, $\ket{10}$ and so on and hence solve the corresponding matrix equations.
\section{}\label{F}
The $kl$-th column of the density matrix with post-selection $\ket{kl}$ can be obtained using Eq. (\ref{37}) for a set of  arbitrary product observables  $C^{(n)}_{A}\otimes C^{(n)}_{B}, n=1,\dots,d$ 
\begin{widetext}
{\scriptsize{
\begin{equation} \label{ F1}
\renewcommand\arraystretch{2.3}
p(\rho_{AB},kl)
    \begin{pmatrix}
    \braket{(C^{(1)}_{A}\otimes C^{(1)}_{B})_w}^{kl}_{\rho_{AB}}\\
    \braket{(C^{(2)}_{A}\otimes C^{(2)}_{B})_w}^{kl}_{\rho_{AB}}\\
    \vdots\\
    \braket{(C^{(d)}_{A}\otimes C^{(d)}_{B})_w}^{kl}_{\rho_{AB}}\\
    \end{pmatrix}
    = \begin{pmatrix}
    [C^{(1)}_{A}]_{k0}[C^{(1)}_{B}]_{l0} & \hspace{2mm} [C^{(1)}_{A}]_{k0}[C^{(1)}_{B}]_{l1} &\cdots& [C^{(1)}_{A}]_{k(d-1)}[C^{(1)}_{B}]_{l(d-1)}\\
    [C^{(2)}_{A}]_{k0}[C^{(2)}_{B}]_{l0} & \hspace{2mm} [C^{(2)}_{A}]_{k0}[C^{(2)}_{B}]_{l1} &\cdots& [C^{(2)}_{A}]_{k(d-1)}[C^{(2)}_{B}]_{l(d-1)}\\
     \vdots&&\ddots\\
   [C^{(d)}_{A}]_{k0}[C^{(d)}_{B}]_{l0} & \hspace{2mm} [C^{(d)}_{A}]_{k0}[C^{(d)}_{B}]_{l1} &\cdots& [C^{(d)}_{A}]_{k(d-1)}[C^{(d)}_{B}]_{l(d-1)}\\
    \end{pmatrix}
     \begin{pmatrix}
   [\rho_{AB}]_{00,kl}\\
    [\rho_{AB}]_{01,kl}\\
    \vdots\\
   [\rho_{AB}]_{(d-1)(d-1),kl}\\
    \end{pmatrix}
\end{equation}
}}
\end{widetext}
With $C^{(1)}_{A}\otimes C^{(1)}_{B}=I_A\otimes I_B$, an identity operator in the joint Hilbert space.  Solution exists for non zero determinant of the above square matrix. The $k'l'$-th column can be calculated measuring the same set of observables with post-selection $\ket{{k{'}l{'}}}$ and by solving the corresponding matrix equations. Remember that post-selections are complete set of projection operators and hence can be measured simultaneously for each product  observables $C^{(n)}_{A}\otimes C^{(n)}_{B}$. So measuring an observable $C^{(n)}_{A}\otimes C^{(n)}_{B}$ according to AAV method, we obtain all the weak values for a complete set of post selections $\{\ket{kl}\}$. One does not need to perform a measurement for the last column as it can be obtained from the normalization and Hermiticity condition of the density matrix.
\section{}\label{G}
The choices for which the quantity LHS - RHS is positive when $p> \frac{3(d-1)}{2(2d-1)}$ are 
\begin{align}
|\hspace{-1mm}\braket{j^{\prime}_B|i^{\prime}_A}\hspace{-1mm}|=1,\hspace{3mm} |\hspace{-1mm}\braket{j^{\prime}_A|i^{\prime}_B}\hspace{-1mm}|=1.\label{G1}
\end{align}
To realize these conditions  physically, we consider spin systems of spin-1 (d=3) and spin-3/2 (d=4) only (one can extend such realizations for higher dimensions also).\\
\textit{(i) Spin-1 system:} the dimension is d=3 with the  basis states  $\ket{0}=(1,0,0)^T, \ket{1}=(0,1,0)^T, \ket{2}=(0,0,1)^T$. The spin operators in x, y and z directions are  $S_x$, $S_y$,  $S_z$ respectively and the ladder operators are $S_{\pm}=S_x{\pm}iS_y$ 
\begin{align*}
 S_x=\frac{1}{\sqrt{2}}&\begin{pmatrix}
   0&1&0\\
    1&0&1\\
   0&1&0
    \end{pmatrix}, 
    S_y=\frac{i}{\sqrt{2}}\begin{pmatrix}
   0&-1&0\\
    1&0&-1\\
   0&1&0
    \end{pmatrix},\\
     &S_z=\begin{pmatrix}
   1&0&0\\
    0&0&0\\
   0&0&-1
    \end{pmatrix},\\
    S_+={\sqrt{2}}&\begin{pmatrix}
   0&1&0\\
    0&0&1\\
   0&0&0
    \end{pmatrix}, 
    S_-={\sqrt{2}}\begin{pmatrix}
   0&0&0\\
    1&0&0\\
   0&1&0
    \end{pmatrix}\nonumber
\end{align*}
 We consider 
\begin{align}
\bra{i^{\prime}_A}C_A=\bra{2}S_x=\bra{1}=\bra{j^{\prime}_A}, \nonumber\\
 \bra{i^{\prime}_B}C_B=\bra{1}S_+=\bra{2}=\bra{j^{\prime}_B}\label{G2}
\end{align}
and hence the condition (\ref{G1}) is fulfilled. The operator $C_A\otimes C_B=S_x\otimes S_+=S_x\otimes S_x + iS_x\otimes S_y$  in the  LHS of  Eq. (\ref{39}) can be realized via product weak values as $\braket{\phi_A\phi_B|(S_x\otimes S_x + iS_x\otimes S_y)\rho|\phi_A\phi_B}=\braket{\phi_A\phi_B|(S_x\otimes S_x)\rho|\phi_A\phi_B} + i\braket{\phi_A\phi_B|(S_x\otimes S_y)\rho|\phi_A\phi_B}$.\\
\textit{Alternative:} 
Consider 
\begin{align}
\bra{i^{\prime}_A}C_A=(\bra{0}+\bra{2})S_x=\bra{1}=\bra{j^{\prime}_A}, \nonumber\\
 \bra{i^{\prime}_B}C_B=\bra{1}S_x=(\bra{0}+\bra{2})=\bra{j^{\prime}_B}\label{G3}
\end{align}
and the condition (\ref{G1}) is satisfied.\\
\textit{(ii) Spin-3/2 system:} the dimension is d=4 with the  basis states  $\ket{0}=(1,0,0,0)^T, \ket{1}=(0,1,0,0)^T, \ket{2}=(0,0,1,0)^T,  \ket{3}=(0,0,0,1)^T$. The spin operators in x, y and z directions are  $J_x$, $J_y$,  $J_z$ respectively and the ladder operators are $J_+$ and  $J_-$
\begin{align}
 J_x&=\frac{1}{{2}}\begin{pmatrix}
   0&\sqrt{3}&0&0\\
    \sqrt{3}&0&2&0\\
   0&2&0&\sqrt{3}\\
   0&0&\sqrt{3}&0
    \end{pmatrix}, \nonumber\\
   J_y&=\frac{i}{{2}}\begin{pmatrix}
   0&-\sqrt{3}&0&0\\
    \sqrt{3}&0&-2&0\\
   0&2&0&-\sqrt{3}\\
   0&0&\sqrt{3}&0
    \end{pmatrix},\nonumber\\
     J_z&=\frac{1}{{2}}\begin{pmatrix}
   3&0&0&0\\
    0&1&0&0\\
   0&0&-1&0\\
   0&0&0&-3
    \end{pmatrix}, \nonumber\\
   J_+&=\begin{pmatrix}
   0&\sqrt{3}&0&0\\
    0&0&2&0\\
   0&0&0&\sqrt{3}\\
   0&0&0&0
    \end{pmatrix},\nonumber\\
    J_-&=\begin{pmatrix}
   0&0&0&0\\
    \sqrt{3}&0&0&0\\
   0&2&0&0\\
   0&0&\sqrt{3}&0
    \end{pmatrix}.\nonumber
\end{align}
Consider the following
\begin{align}
\bra{i^{\prime}_A}C_A=\bra{3}J_x=\bra{2}=\bra{j^{\prime}_A}, \nonumber\\
 \bra{i^{\prime}_B}C_B=\bra{2}J_+=\bra{3}=\bra{j^{\prime}_B}\label{G4}
\end{align}
and  the condition (\ref{G1}) is fulfilled. Note that, normalizing factors in the calculations (\ref{G2}), (\ref{G3}) and (\ref{G4}) are not important as these factors will get canceled out in Eq. (\ref{39}).  
\end{document}